\documentclass[apj]{emulateapj}
\usepackage{amsmath}
\usepackage{apjfonts}

\newcommand{\DDir}{\relax{D\kern-.7em{/}}}

\newcommand{\haf}{\frac{1}{2}}
\newcommand{\inv}[1]{\frac{1}{#1}}

\newcommand{\ra}{\rightarrow}

\newcommand{\xra}{\xrightarrow}

\newcommand{\X}{\times}

\newcommand{\be}{\begin{equation}}
\newcommand{\ee}{\end{equation}}
\newcommand{\bea}{\begin{equation*}}
\newcommand{\eea}{\end{equation*}}

\newcommand{\abs}[1]{\left\vert#1\right\vert}

\newcommand{\ave}[1]{\left\langle #1\right\rangle}

\newcommand{\nin}{\relax{\in\kern-.8em{/}}}

\newcommand{\te}{\theta}
\newcommand{\al}{\alpha}
\newcommand{\bt}{\beta}

\newcommand{\lm}{\lambda}
\newcommand{\Lm}{\Lambda}

\newcommand{\De}{\Delta}
\newcommand{\Ga}{\Gamma}

\newcommand{\sig}{\sigma}

\newcommand{\vep}{\varepsilon}
\newcommand{\ep}{\epsilon}
\newcommand{\zt}{\zeta}

\newcommand{\vs}{\textrm{v}_s}

\newcommand{\cm}{\mbox{ cm}}

\newcommand{\se}{\mbox{ s}}
\newcommand{\yr}{\mbox{ yr}}
\newcommand{\kyr}{\mbox{ kyr}}
\newcommand{\erg}{\mbox{ erg}}
\newcommand{\Hz}{\mbox{ Hz}}

\newcommand{\GHz}{\mbox{ GHz}}

\newcommand{\km}{\mbox{ km}}
\newcommand{\pc}{\mbox{ pc}}
\newcommand{\kpc}{\mbox{ kpc}}

\newcommand{\eV}{\mbox{ eV}}
\newcommand{\keV}{\mbox{ keV}}
\newcommand{\MeV}{\mbox{ MeV}}
\newcommand{\GeV}{\mbox{ GeV}}
\newcommand{\TeV}{\mbox{ TeV}}

\newcommand{\muG}{\mbox{ $\mu$G}}

\newcommand{\sref}{\S~\ref}

\begin{document}
\newcommand{\vv}{\textrm{v}}
\newcommand{\SNRa}{RX J1713.7-3946}
\newcommand{\SNRb}{RX J0852.0-4622}
\newcommand{\nuTeV}{\nu_{\text{TeV}}}
\newcommand{\nukeV}{\nu_{\text{keV}}}
\newcommand{\nuGHz}{\nu_{\text{GHz}}}
\newcommand{\tkyr}{t_{\text{kyr}}}
\newcommand{\dkpc}{d_{\text{kpc}}}
\newcommand{\IC}{\text{IC}}
\newcommand{\PP}{\text{PP}}
\newcommand{\TB}{\text{TB}}
\newcommand{\ST}{\text{ST}}
\newcommand{\Syn}{\text{Syn}}
\title{In which shell-type SNRs should we look for gamma-rays and neutrinos from p-p collisions?
}
\author{Boaz Katz\altaffilmark{1} and Eli
Waxman\altaffilmark{1}}

\altaffiltext{1}{Physics Faculty, Weizmann Institute, Rehovot 76100, Israel; boazka@wizemail.weizmann.ac.il, waxman@wicc.weizmann.ac.il}

\begin{abstract}
We present a simple analytic model for the various contributions to the non-thermal emission from shell type SNRs, and show that this model's results reproduce well the results of previous detailed calculations.
We show that the $\geq 1\TeV$ gamma ray emission from the shell type SNRs \SNRa{} and \SNRb{} is dominated by inverse-Compton scattering of CMB photons (and possibly infra-red ambient photons) by accelerated electrons.
Pion decay (due to proton-proton collisions) is shown to account for only a small fraction, $\lesssim10^{-2}$, of the observed flux, as assuming a larger fractional contribution would imply nonthermal radio and X-ray synchrotron emission and thermal X-ray Bremsstrahlung emission that far exceed the observed radio and X-ray fluxes. Models where pion decay dominates the $\geq 1\TeV$ flux avoid the implied excessive synchrotron emission (but not the implied excessive thermal X-ray Bremsstrahlung emission) by assuming an extremely low efficiency of electron acceleration, $K_{ep}\lesssim 10^{-4}$ ($K_{ep}$ is the ratio of the number of accelerated electrons and the number of accelerated protons at a given energy).
We argue that observations of SNRs in nearby galaxies imply a lower limit of $K_{ep}\gtrsim 10^{-3}$, and thus rule out $K_{ep}$ values $\lesssim 10^{-4}$
(assuming that SNRs share a common typical value of $K_{ep}$).
It is suggested that SNRs with strong thermal X-ray emission, rather than strong non-thermal X-ray emission, are more suitable candidates for searches of gamma rays and neutrinos resulting from proton-proton collisions. In particular, it is shown that the neutrino flux from the SNRs above is probably too low to be detected by current and planned neutrino observatories.
Finally, we note that the magnetic field value implied by the comparison of X-ray to gamma-ray emission, $\sim10\muG$, can be used to constrain magnetic field amplification.
\end{abstract}

\section{Introduction}\label{sec:Introduction}
For a long time it is believed that the galactic cosmic rays observed up to the 'knee' energy ($\sim10^{15} \eV$) are accelerated in supernova remnants \citep[SNRs, see e.g.][]{Axford94}.
The relativistic protons (and electrons) are believed to be accelerated by the diffusive (Fermi) shock acceleration (DSA) mechanism (for reviews see \citealt{Drury83,Blandford87,Malkov01}).
Strong evidence for electron acceleration to high energies in SNRs was established by observations of non-thermal X-ray emission which was attributed to synchrotron radiation of multi $\TeV$ electrons \citep{Koyama95}, while direct evidence for ion acceleration in SNRs has not been presented so far.

Recently, unambiguous detection of $\gtrsim1\TeV$ $\gamma$-rays has been made from the shell-type SNRs \SNRa{} \citep{Muraishi00,Enomoto02,Aharonian04,Aharonian06} and \SNRb{} \citep{Katagiri05,Aharonian05,Aharonian07}, providing the first direct proof for the acceleration of particles at SNRs to multi-\TeV{} energies.
There are two candidate emission processes that can account for this radiation, namely inverse Compton (IC) of radio and infra red photons by multi \TeV{} accelerated electrons or pion decay as a consequence of proton-proton (PP) interactions of multi \TeV{} accelerated protons with ambient target protons \citep[e.g.][]{Drury94}.

The distinction between the two mechanisms has important consequences for understanding particle acceleration and magnetic field amplification in SNRs. If it would turn out that the source is PP emission, this would be the first direct evidence for proton acceleration in SNRs.
An IC source would allow a rather accurate estimate of the downstream magnetic field value by comparing the X-ray to gamma ray fluxes \citep[see e.g. ][]{Aharonian06}.

Broad-band emission models with different levels of sophistication where applied in order to analyze the observed non-thermal radiation from these SNRs, reaching different conclusions as to the dominant $\gtrsim 1\TeV$ $\gamma$-ray emission mechanism.
For \SNRa{} \citet{Aharonian06}, \citet{Berezhko06} and \citet{Moraitis07} claimed that PP emission is favorable and IC is unlikely, \citet{Porter06} claimed that IC emission is consistent.
For \SNRb{} \citet{Enomoto06} and \citet{Aharonian07} did not rule out either mechanism.

The only well understood non-thermal emission mechanism is currently Synchrotron radiation of accelerated electrons. Synchrotron radiation is primarily observed in radio frequencies and is observed in X-rays in a few known SNRs \citep[see e.g. ][and refferences within]{Bamba05}. Non-thermal radio and X-ray observations are crucial for studying the non-thermal electron population. In the few known examples of SNRs emitting X-ray Synchrotron radiation, the X-ray flux (per logarithmic frequency) is decreasing, indicating that the flux peaks at lower, unresolved photon energies. The radio and X-ray luminosities, and the implied position of the cutoff in the X-ray spectrum, may be used (and have been used in the models discussed above) for constraining the accelerated electron distribution, and thus for constraining the expected IC emission.

In this paper we derive simple analytic relations between the dominant radio, X-ray and $\gamma$-ray emission mechanisms that can be used for distinguishing between IC and PP origins of the $\gamma$-rays in SNRs were $\gamma$-rays were observed, and for predicting the $\gamma$-ray and neutrino flux values for SNRs where only radio and/or X-ray emissions where detected.
Whenever possible, the simple analytic approximations we find are compared to, and shown to agree with, previous detailed calculations, with the advantage of being easier to follow and maintaining the explicit dependence on the unknown parameters.

First we compare in section \sref{sec:OneZone} the expected IC, PP, Synchrotron and Thermal Bremsstrahlung (TB) fluxes in a simple one zone model of shocked ISM plasma. The plasma is assumed to consist of thermal and accelerated electron and proton components with the later consisting of relativistic particles having a power law distribution in energy. In this section, the high energy cutoffs of the accelerated electron and proton distributions are ignored.
Next, we discuss in section \sref{sec:EnergyCutoffs} the maximal energy attainable by electrons and protons in a SNR due to cooling and limited SNR age, and the implications for the non-thermal emitted spectra, assuming diffusive shock acceleration (DSA) as the acceleration mechanism.
We then find in section \sref{sec:M33} an upper limit to the value of $K_{ep}$, the ratio of the number of accelerated electrons and the number of accelerated protons at a given energy, $K_{ep}>10^{-3}$ by studying the radio observations of SNRs in M33. This parameter enters into the ratios of IC and Synchrotron emission to PP emission. This lower limit is used to rule out previously suggested SNR broadband emission models that used considerably lower values.
In section \sref{sec:SNRSab} we apply the results of earlier sections to show that the broad-band spectrum of the SNRs \SNRa{} and \SNRb{} is inconsistent with a PP origin and is consistent with an IC origin of the $\gtrsim 1\TeV$ emission. We discuss previous claims that this emission cannot be due to IC and argue against them.
The results are summarized and discussed in \sref{sec:Discussion}.

\section{PP emission Vs. Thermal and non thermal electronic emission}\label{sec:OneZone}
In this section we compare thermal and non thermal continuum emission mechanisms in the shocked plasma behind SNR blastwaves.
The non thermal emission is assumed to be emitted by relativistic, accelerated electrons and protons with power law distributions in energy.
In this section we ignore the energy cutoffs of the accelerated particle distributions. This issue is discussed in \sref{sec:EnergyCutoffs}.
We focus on ratios of the expected fluxes which are weakly dependent on unknown parameters such as distance to the remnant and total energy.

First we write down  in \sref{sec:Emissions} simple expressions for the luminosities due to the different processes in simple forms that allow easy comparison with each other (a derivation of these equations is given in \sref{sec:EmmisionsA}).
Next, we compare in \sref{sec:ICPP} the two $\sim 1\TeV$ $\gamma$-ray emission mechanisms, IC and PP.
We then derive in \sref{sec:PPICTBSyn} constraints on $\gamma$-ray PP and IC emission by comparing them to thermal X-ray Bremsstrahlung and to radio synchrotron emission.
Finally, the results of \sref{sec:Emissions}-\sref{sec:PPICTBSyn} are compared in \sref{sec:Comparison1} to earlier studies of the SNRs \SNRa{} and \SNRb{} (the only shell-type SNRs that are known to emit $\gtrsim1\TeV$ $\gamma$-rays).

We note that most of the results presented in this section are not restricted to SNRs and are applicable to any system that efficiently accelerates protons and/or electrons to relativistic energies with power-law energy distributions.

\subsection{Emission mechanisms}\label{sec:Emissions}
Consider the shocked plasma in the downstream of the blastwave of a SNR. Here we consider radiation emitted by four distinct particle populations:
\begin{enumerate}
\item Thermal electron and proton components with similar number densities, $n_e\sim n_p\equiv n$, which we assume consist of most of the particles. For simplicity we assume that the electron and proton energy distributions are given by Maxwelians with temperatures $T_e$ and $T_p$ respectively with $T_e=\zt_eT_p$.
The total number of protons or electrons is $N$ and the total thermal energy is $E_{\text{th}}\approx(3/2)NT_p$.
\item Power law distributions of relativistic accelerated electrons and protons with an electron:proton ratio $K_{ep}$,
\begin{align}\label{eq:Powerlaw}
&\frac{dN_e}{d\vep_e}|_{\vep_e=\vep_p}=K_{ep}\frac{dN_p}{d\vep_p}=K_{ep}\frac{E_p}{\vep_{p,\min}^2\Lm_p}\left(\frac{\vep_p}{\vep_{p,\min}}\right)^{-p},
\end{align}
where $\vep_e,\vep_p$ are the electron and proton energies respectively, $p$ is the power law index assumed to be $p\approx 2$, $E_p$ is the total energy in accelerated protons and
\begin{align}\label{eq:Lambdap}
&\Lm_p\approx\inv{p-2}\left[1-\left(\frac{\vep_{p,\max}}{\vep_{p,\min}}\right)^{-(p-2)}\right]\cr &\xra[p\ra2]{}\log\left(\frac{\vep_{p,\max}}{\vep_{p,\min}}\right).
\end{align}
The distribution of the protons is described by \eqref{eq:Powerlaw} for proton energies $\vep_{p,\min}<\vep_p<\vep_{p,max}$ with $\vep_{p,min}\sim m_pc^2$. The value of $\vep_{\max}$ depends on the SNR parameters and acceleration mechanism. Estimates of $\vep_{\max}$ assuming DSA will be derived in \sref{sec:EnergyCutoffs}.
\end{enumerate}

We study the following radiation emission mechanisms:
\begin{enumerate}
\item $\gamma$-rays and neutrinos emitted as a result of proton-proton collisions (PP) between the relativistic protons and the thermal protons.
The PP gamma-ray luminosity per logarithmic photon energy is given by [cf. Eq. \eqref{eq:PPExactA}]:
\begin{align}\label{eq:PPExact}
&\nu L_{\nu~\PP}=C_{\PP}(p)2\vep_{p}(\nu)\frac{dN_p}{d\vep_p}\sig^{\text{inel}}_{pp} n c h\nu,
\end{align}
where $\vep_p dN_p/d\vep_p$ is to be evaluated at $\vep_{p,\PP}(\nu)=10h\nu$ (photon energies are referred to through the photon frequency throughout the paper), the typical proton energy for which photons with energy $h\nu$ are emitted. For $p=2,2.2$ we have $C_{\PP}(2)\approx 0.85, C_{\PP}(2.2)\approx 0.66$.
The neutrino luminosity is similar to the $\gamma$-ray luminosity at equal photon and neutrino energies.
\item $\gamma$-rays emitted by Inverse Compton (IC) resulting from the interaction of the relativistic electrons with CMB photons.
The IC gamma-ray luminosity per logarithmic photon energy is given by [cf. Eq. \eqref{eq:ICExactA}]:
\begin{align}\label{eq:ICExact}
\nu L_{\nu~\IC}=C_{\IC}(p)\haf \vep_e\frac{dN_e}{d\vep_e}\frac43 \sig_T\gamma_e^2(\nu)U_{\text{CMB}}c,
\end{align}
where $T_{\text{CMB}},U_{\text{CMB}}=aT_{\text{CMB}}^4$ are the temperature and energy density of the CMB photons.
$\vep_e dN_e/d\vep_e$ is to be evaluated at $\vep_{e,IC}(\nu)=\gamma_e(\nu)m_ec^2\equiv m_ec^2(h\nu/3T_{\text{CMB}})^{1/2}$, the typical electron energy for which electrons up scatter CMB photons to energy $h\nu$.
The correction factor, $C_{\IC}(p)$, is approximately $C_{\IC}(p)\approx 0.8$ (to within $5\%$) for $2<p<2.2$.
It is useful to note that $\gamma_e^2(\nu)U_{\text{CMB}}=[U_{\text{CMB}}/(3T_{\text{CMB}})]h\nu\approx 0.9 n_{\text{CMB}} h\nu$ where $n_{\text{CMB}}$ is the number density of CMB photons.

\item Radio and X-ray synchrotron (Syn) emission of the relativistic electrons in an assumed magnetic field $B$.
The synchrotron luminosity per logarithmic frequency is given by [cf. Eq. \eqref{eq:SynExactA}]:
\begin{align}\label{eq:SynExact}
&\nu L_\nu^{\Syn}=C_{\Syn}(p)\haf \vep_e(\nu)\frac{dN_e}{d\vep_e}\frac43\sig_T\gamma_e^2(\nu)U_Bc
\end{align}
where $U_{B}=B^2/(8\pi)$.
$\vep_e dN_e/d\vep_e$ is to be evaluated at $\vep_e(\nu)=\gamma_e(\nu)m_ec^2\equiv (2\nu/\nu_B)^{1/2}m_ec^2$, the typical energy of electrons emitting photons with frequency $\nu$,
where $\nu_{B}\equiv qB/(2\pi m_ec)$.
The correction factor, $C_{\Syn}(p)$, is approximately $C_{\Syn}(p)\approx 0.8$ (to within $5\%$) for $2\leq p<2.2$.
\item Thermal-Bremsstrahlung (TB) emission of the thermal electrons interacting with the thermal protons.
The maximal TB luminosity per logarithmic frequency is emitted at the photon energy $h\nu=T_e$ and is given by [cf. Eq. \eqref{eq:TBExactA}]:
\begin{align}\label{eq:TBExact}
&\nu L_{\nu,h\nu=T_e}^{\TB}=\sqrt{\frac{8}{3\pi}}e^{-1}\al_e\bar g_{ff}N\sig_Tnc\sqrt{m_ec^2T_e}
\end{align}
where $e$ is the natural logarithm, $\al_e\approx1/137$ is the fine structure constant and $\bar g_{ff}$ is the thermal Gaunt factor.
For $100\eV<T_e<10\keV$, the value of $\bar g_{ff}$ (for $h\nu=T_e$) is in the range, $0.8<\bar g_{ff}<1.2$ \citep[e.g. ][]{Karzas61}.
\end{enumerate}

We note that the amount of secondary electrons and positrons resulting from PP interactions is most likely negligible compared to the primary population of accelerated electrons. The energy output in electrons and positrons per logarithmic particle energies is roughly equal to the $\gamma$-ray emission given by Eq. \eqref{eq:PPExact}. The ratio of secondary electrons+positrons to protons for an SNR of age $t=1000t_{\kyr}\yr$ evolving into a medium with proton density $n=n_0 \cm^{-3}$ is thus roughly given by (ignoring cooling, which affects both primary and secondary populations in the same way)
$\vep^2dN_{e+e-}/d\vep\sim 0.2 \vep^2dN_p/d\vep_p\sig^{\text{inel}}_{pp} n c t\sim 10^{-6}\vep^2dN_p/d\vep_p t_{\kyr}n_0$. As long as $K_{ep}\gg10^{-6}t_{\kyr}n_0$, the contribution of the secondary electrons to the broad band emission is negligible. Henceforth we ignore this contribution.

\subsection{IC to PP emission ratio}\label{sec:ICPP}
Here we directly compare the two competing TeV $\gamma$-ray emission mechanisms.
Ignoring the possible cutoffs of the spectrum of both species, the ratio of expected IC to PP gamma ray luminosities per photon frequency can be approximated by [compare Eq. \eqref{eq:PPExact} and Eq. \eqref{eq:ICExact}]:

\begin{align}\label{eq:ICPP}
&L_{\nu~\IC}/L_{\nu~\PP}\approx0.3
\frac{\vep_{e,\IC}(\nu)dN_e/d\vep_e}{\vep_p(\nu)dN_p/d\vep_p}\frac{\sig_T}{\sig^{\text{inel}}_{pp}}\frac{n_{\text{CMB}}}{n}\cr
&\approx 10K_{ep,-2}\nuTeV^{(p-1)/2}n_0^{-1},
\end{align}
where $K_{ep}=10^{-2}K_{ep,-2}$ and $n=1n_0\cm^{-3}$. It is useful to note that
\begin{align}
&\frac{\vep_{e,\IC}(\nu)dN_e/d\vep_e}{\vep_p(\nu)dN_p/d\vep_p}=K_{ep}\left[\frac{\vep_{e,\IC}(\nu)}{\vep_p(\nu)}\right]^{-(p-1)}
\end{align}
and
\begin{align}
&\frac{\vep_{e,\IC}(\nu)}{\vep_p(\nu)}=\inv{10}\frac{m_ec^2}{\sqrt{3T_{\text{CMB}}h\nu}}\approx 2 \nuTeV^{-1/2}.
\end{align}
Comparison of Eq. \eqref{eq:ICPP} with the results of previous studies is presented in \sref{sec:Comparison1}.

An electron to proton ratio of order $K_{ep}\sim 10^{-2}$ is commonly assumed based on the measured electron:proton ratio in the cosmic rays \citep[see e.g. ][]{Longair94} under the assumption that SNRs are the main source of proton and electron cosmic rays.
In section \sref{sec:M33} we find a lower limit of $K_{ep}\gtrsim 10^{-3}$ based on radio observations of SNRs in M33.

Using Eq. \eqref{eq:ICPP}, we see that as long as electron cooling does not suppress the IC flux, IC dominates PP emission as long as
\begin{align}\label{eq:MinnPPIC}
n\lesssim 10K_{ep,-2}\nuTeV^{(p-1)/2}\cm^{-3}.
\end{align}

The effect of electron cooling is addressed in \sref{sec:EnergyCutoffs}.

\subsection{PP and IC to TB X-rays and non thermal radio Syn}\label{sec:PPICTBSyn}
Here we compare the expected PP and IC $\gamma$-ray emission to thermal and synchrotron emission. This is useful for constraining the expected gamma-ray and neutrino fluxes based on observed radio and X-ray fluxes.

By comparing equation \eqref{eq:PPExact} to \eqref{eq:TBExact} we see that the ratio of PP $\gamma$-ray luminosity at photon energies $h\nu_{\gamma}$ to the X-ray TB luminosity at photon energies $h\nu_X=T_e$ (the photon energy of maximal emission per logarithmic photon energy) can be written as:
\begin{align}\label{eq:PPTB1}
&\frac{\nu_{\gamma} L_{\nu_{\gamma}~\PP}}{\nu_{X}L_{\nu_{X}~\TB}|_{h\nu=T_e}}=\cr
&=\frac{3e}{10}\sqrt{\frac{3\pi}{8}}\al_{e}^{-1}\frac{C_{pp}(p)}{\bar g_{ff}}\frac{\vep_p^2(\nu_{\gamma})dN_p/d\vep_p}{E_{\text{th}}}\frac{T_p}{\sqrt{m_ec^2T_e}}\frac{\sig^{\text{inel}}_{pp}}{\sig_T}
\end{align}
where $E_{\text{th}}\approx(3/2)NT_p$ is the total thermal energy.
Using equation \eqref{eq:Powerlaw} this can be written as:
\begin{align}\label{eq:PPTB2}
&\frac{\nu_{\gamma} L_{\nu_{\gamma}~\PP}}{\nu_{X}L_{\nu_{X}~\TB}|_{h\nu=\zt_eT_p}}\approx 3\X10^{-3}\ep_{p,-1}\zt_{e}^{-1/2}T_{p,\text{keV}}^{1/2}\cr
&\X\Lm_{p,1}^{-1}(10^3\nuTeV)^{-(p-2)}
\end{align}
where $\ep_p=0.1\ep_{p,-1}=E_p/E_{\text{th}}$, $\Lm_p=10\Lm_{p,1}$, $T_p=T_{p,\text{keV}}\keV$, $h\nu_{\gamma}=\nuTeV\TeV$ and we substituted $\vep_{\min}\sim m_pc^2$.
Assuming $p\geq 2$, $\vep_{\max}>10\TeV$ and $h\nu_{\gamma}>10\GeV$ the factor in the second line of Eq.~\eqref{eq:PPTB2} is smaller than $1.1$.

Temperatures $T_p\gtrsim\keV$ are characteristic of the shocked plasma in young SNRs with blastwaves propagating at velocities $\vs\gtrsim 1000\km\se^{-1}$.
In fact, the proton temperature behind a strong shock propagating with velocity $\vs$ is given by:
\begin{align}\label{eq:Tp}
&T_p=\frac3{16}m_p\vs^2=2\vv_{8}^2 \keV,
\end{align}
where $\vs=1000\vv_{8}\km\se^{-1}$ and an adiabatic index equal to $\gamma=5/3$ was assumed.
A lower limit to the shock velocity and thus to $T_p$ for SNRs where non-thermal X-rays are observed is discussed in \sref{sec:EnergyCutoffs}.

The ratio $\zt_e$ of electron to proton temperatures depends on the amount of collisionless heating in the shock and the following heating through Coulomb scattering. The amount of collissionless heating for high Mach shocks is not really known \citep[for a recent review see ][]{Rakowski05}. A lower limit to $\zt_e$ can be derived by assuming that there is no collisionless heating.
After a time $t=t_{\text{kyr}}\kyr$ the ratio would be
[cf. Eq. \eqref{eq:TeTpCoulomba}]:
\begin{align}\label{eq:TeTpCoulomb}
&\zt_e\gtrsim0.6\left(\lm_{ep,1.5}n_0t_{\text{kyr}}
\right)^{2/5}T_{p,\text{keV}}^{-3/5},
\end{align}
where $\lm_{ep}=30\lm_{ep,1.5}$ is the Coulomb logarithm. Eq. \eqref{eq:TeTpCoulomb} was derived assuming $m_e/m_p\ll\zt_e\ll1$ and is valid as long as the resultant value is in this range.
Substituting Eq. \eqref{eq:TeTpCoulomb} in Eq. \eqref{eq:PPTB2} we find:

\begin{align}\label{eq:PPTBCoulomb}
&\frac{\nu_{\gamma} L_{\nu_{\gamma}~\PP}}{\nu_{X}L_{\nu_{X}~\TB}|_{h\nu=T_e}}\lesssim
4\X10^{-3}
\ep_{p,-1}T_{p,\text{keV}}^{4/5} \left(\lm_{ep,1.5}n_0t_{\text{kyr}}
\right)^{-1/5}\cr &\X\Lm_{p,1}^{-1}(10^3\nuTeV)^{-(p-2)}.
\end{align}

By comparing equations \eqref{eq:PPExact} and \eqref{eq:SynExact} we see that the expected ratio of $\gamma$-ray PP luminosity at photon energies $h\nu_{\gamma}$ to the radio synchrotron luminosity at frequency $\nu_R$ can be approximated by:
\begin{align}\label{eq:PPSynGHz}
&\frac{\nu_{\gamma} L_{\nu_{\gamma}~\PP}}{\nu_{R}L_{\nu_{R}~\Syn}}\approx3
\frac{\vep_p(\nu_\gamma)dN_p/d\vep_p}{\vep_{e,\Syn}(\nu_R)dN_e/d\vep_e}\frac{\sig^{\text{inel}}_{pp}}{\sig_T}\frac{nh\nu_{\gamma}}{\gamma_{e,\Syn}(\nu_R)^2U_B}\cr
&\approx 50K_{ep,-2}^{-1}B_{-5}^{-3/2}n_0\nuGHz^{-1/2}\X\cr
&~~~~~\left(2\X10^3B_{-5}^{1/2}\nuGHz^{-1/2}\nuTeV\right)^{-(p-2)}
\end{align}
where $\gamma_{e,\Syn}(\nu_R)=\vep_{e,\Syn}(\nu_R)/m_ec^2=(4\pi m_ec\nu_R/qB)^{1/2}$ is the typical gamma factor of electrons emitting radiation with \text{$\nu_R=\nuGHz\GHz$} frequency and $B=10B_{-5}\muG$.
Assuming $p\geq 2$, the factor in the second line of Eq.~\eqref{eq:PPSynGHz} is smaller or equal to 1.
It is useful to note that,
\begin{align}
&\frac{\vep_p(\nu_\gamma)}{\vep_{e,\Syn}(\nu_R)}=\frac{10\sqrt{h\nu h\nu_B/2}}{m_ec^2}\approx 2\X10^3 B_{-5}^{1/2}\nuGHz^{-1/2}\nuTeV.
\end{align}
Comparison of Eq. \eqref{eq:PPSynGHz} with the results of previous studies is presented in \sref{sec:Comparison1}.

Ignoring the possible cutoff of the IC spectrum, the expected ratio of $\TeV$ IC emission to $\GHz$ synchrotron emission is approximately given by [compare Eq. \eqref{eq:ICExact} and Eq. \eqref{eq:SynExact}]:
\begin{align}\label{eq:ICSynGHz}
&\frac{\nu_{\IC}L_{\nu_{\IC}}}{\nu_{\Syn}L_{\nu_{\Syn}}}\approx \left(\frac{\vep_{e,\IC}(\nu_{\IC})}{\vep_{e,\Syn}(\nu_{\Syn})}\right)^{3-p}\frac{U_{\text{CMB}}}{U_B}\sim 500 B_{-5}^{-3/2}\nuTeV^{1/2}\nuGHz^{-1/2}\cr
&\X\left(4\X10^3 \nuTeV^{1/2}\nuGHz^{-1/2}B_{-5}^{1/2}\right)^{-(p-2)}.
\end{align}

The luminosity ratios of the different radio, X-ray and $\gamma$-ray emission mechanisms are given by equations \eqref{eq:ICPP},\eqref{eq:PPTB2}, \eqref{eq:PPSynGHz} and \eqref{eq:ICSynGHz}.
Flux normalization is obtained by noting that the expected radio flux per logarithmic frequency for a SNR with total energy $E=10^{51}E_{51}\erg$ and a fraction $\eta_p=0.1\eta_{p,-1}$ of the total energy carried by accelerated protons $E_p=\eta_pE$ ($\eta_p\sim \epsilon_p/2$) located at a distance $d=\dkpc\kpc$, is approximately given by [using Eq.~\eqref{eq:SynExact}],
\begin{align}\label{eq:SynFlux}
&\nu f_{\nu,\Syn}|_{\nu=1\text{GHz}}\sim 4\X10^{-13}K_{ep,-2}\eta_{p,-1}E_{51}B_{-5}^{3/2}\dkpc^{-2}\erg\cm^{-2}\se^{-1}.
\end{align}

\subsection{Comparison with previous studies}\label{sec:Comparison1}
Next we compare the results presented in this section to previous studies of the broad-band emission of SNRs. We focus on studies that were published following the discovery of the $\gtrsim 1\TeV$ $\gamma$-rays from the shell-type SNRs \SNRa{} and \SNRb{} (the broad-band emission of these SNRs is discussed in \sref{sec:SNRSab}).
Recent broad-band studies of \SNRa{} where done in \citep{Aharonian06,Berezhko06,Porter06,Moraitis06} while studies of \SNRb{} include \citep{Enomoto06,Aharonian07}.
These studies differ in the way the particle distributions are obtained, in the assumptions regarding the magnetic field value and in the assumptions regarding the ambient IR radiation field. \citet{Berezhko06} numerically solved time-dependent CR transport equations, coupled nonlinearly with the hydrodynamic equations for the thermal component.
\citet{Aharonian06}, \citet{Porter06} and \citet{Aharonian07} assumed a constant injection of particles with a power-law spectrum that is cutoff exponentially over a fixed period of time. They calculated numerically the effects of cooling on the particle spectrum \citep[][take in addition particle escape into consideration]{Aharonian07}. \citet{Moraitis06} found an analytic solution to "two-zone" (acceleration zone and escape zone) spatially averaged kinetic equations that include cooling.
\citet{Enomoto06} assume a power-law spectrum that is cutoff exponentially. \citet{Berezhko06} estimated the value of the magnetic field based on observation of thin X-ray filaments (see discussion in \sref{sec:ClaimsNotIC}) while the other authors allowed for different magnetic field values. \citet{Porter06} included a detailed model of the galactic radiation field with IR and CMB dominating in different places, while the other authors assumed 'standard' averaged values.

All the above studies focused on the non-thermal emission mechanisms only.
Comparisons of Eqs. \eqref{eq:ICPP} and \eqref{eq:PPSynGHz} with the results of these studies are presented in tables \ref{table:ICtoPP} and \ref{table:PPtoSyn} respectively.
\begin{table}[ht]
\caption{Ratio of IC to PP emission for negligible suppression of IC due to cooling}
\begin{center}
\begin{tabular}{|l|l|l|l||l|}
\hline
Ref. & $K_{ep}$ & $n[\cm^{-3}]$ & $L_{\nu~\IC}/L_{\nu~\PP}(\TeV)$ &  Eq. \eqref{eq:ICPP}\\
\hline
1& $\approx 5\X10^{-4}\footnotemark[a]$& $1$& $\approx 0.3$ & $0.5$\\
\hline
2a& $1.7\X10^{-3}$& 0.008& $\approx 200$ &$200$\\
2b& $3.5\X10^{-2}$ &0.01&$\approx 7000$&3500\\
2c& $10^{-2}$& 0.2& $\approx 30$& 50\\
\hline
\end{tabular}
\end{center}
The value of $L_{\nu~\IC}/L_{\nu~\PP}(\TeV)$ that results from Eq. \eqref{eq:ICPP} for the values of $K_{ep}$ and $n[\cm^{-3}]$ used in each reference is shown in the last column next to that obtained in each reference (fourth column). References: (1) \citet{Moraitis06}; (2) \citet{Aharonian07}, figure 17a; (3) \citet{Aharonian07}, figure 17b; (3) \citet{Enomoto06};\\
The line distinguishes between PP and IC dominated models of SNRs \SNRa{} and \SNRb{}.
\footnotetext[a]{The value of $K_{ep}$ was calculated by $K_{ep}\approx Q_0p_0^{s_2-1}\bar Q_0^{-1}\bar p_0^{-(s_2-1)}$ assuming the escape zone dominates and that $p,p_0\ll p_{\max}$, using the authors notations.}
\label{table:ICtoPP}
\end{table}
\begin{table}[ht]
\caption{Ratio of PP to synchrotron emission}
\begin{center}
\begin{tabular}{|l|l|l|l|l|l||l|}
\hline
Ref. & $K_{ep}$ & $n[\cm^{-3}]$ & B[\muG] & p & PP/Syn\footnotemark[a] & Eq. \eqref{eq:PPSynGHz}\\
\hline
1& $\approx 10^{-4}$& $1$\footnotemark[b]& 130 & &$\approx 200$ &100(p=2)\\
2& $\approx 5\X10^{-4}\footnotemark[c]$& $1$& 15 &2.07 &$\approx 250$ & 300\\
3a& $2.4\X10^{-6}$& 0.2& 120 & 2.1&$\approx300$ &400\\
3b& $4.5\X10^{-4}$& 2& 85 & 2& $\approx 70$ &100\\
\hline
3c& $1.7\X10^{-3}$& 0.008&6&2.4&$\approx 0.5$ &$0.3$\\
3d& $3.5\X10^{-2}$ &0.01&6.5&2.4&$\approx 0.3$&$0.15$\\
4& $10^{-2}$&0.2&$\approx6$&2.1&$\approx 2$&10\\
\hline
\end{tabular}
\end{center}
The value of $\nu L_{\nu~\PP}(\TeV)/\nu L_{\nu~\Syn}(\GHz)$ that results from Eq. \eqref{eq:PPSynGHz} for the values of $K_{ep}$, $n[\cm^{-3}]$, $B[\muG]$ and $p$ used in each reference is shown in the last column next to that obtained in each reference (sixth column).
References: (1) \citet{Berezhko06}; (2) \citet{Moraitis06}; (3a) \citet{Aharonian07}, figure 18a ;(3b) \citet{Aharonian07}, figure 18b; (3c) \citet{Aharonian07}, figure 17a; (3d) \citet{Aharonian07}, figure 17b; (4) \citet{Enomoto06}\\
The line distinguishes between PP and IC dominated models of SNRs \SNRa{} and \SNRb{}.
\footnotetext[a]{$\nu L_{\nu~\PP}(\TeV)/\nu L_{\nu~\Syn}(\GHz)$}
\footnotetext[b]{In this reference the ambient density is nonuniform, $n$ is taken as the\\ value of the ambient number density currently encountered by the shock}
\footnotetext[c]{See footnote a in table \ref{table:ICtoPP}}
\label{table:PPtoSyn}
\end{table}
As can be seen, there is good agreement (up to a factor $\sim2$) between our analytic expressions, Eqs. \eqref{eq:ICPP} and \eqref{eq:PPSynGHz}, and the results of earlier detailed numerical calculations of the remnants \SNRa{} and \SNRb{}.

\section{Energy cutoffs}\label{sec:EnergyCutoffs}
In section \sref{sec:OneZone} we discussed the radiation emitted by power-law distributed electrons and protons. In reality, the particle distribution functions can be approximated by a power law function only over a limited range of particle energies.
The maximal energies and corresponding cutoff frequencies in the emitted spectrum were extensively studied before \citep[see e.g. ][]{Drury94, Reynolds98}.
For completeness we write down in this section the expressions for the maximal particle energies attainable by DSA in a simple SNR model and the possible spectral cooling break in the electron spectrum and discuss the implications for the spectrum of the emitted radiation.

First we write down in \sref{sec:Cutoffs} the maximal energies attainable by DSA as a function of the SNR radius, age and energy, ignoring the dynamical relation between these quantities.
Next, we focus in \sref{sec:Dynamics} on the Sedov-Taylor (ST) SNR evolution phase.
We then discuss in \sref{sec:NonThermal} SNRs in which non-thermal X-rays are observed. We find a lower limit to the shock velocity and post shock temperature in such SNRs. In addition we derive constraints on the $\gamma$-ray and X-ray spectral cutoffs that must be satisfied by an IC model for the $\gamma$-rays emitted by such SNRs.
Finally, we find in \sref{sec:ICSupp} an upper limit to the PP emission for SNRs in which the IC $\TeV$ emission is suppressed by synchrotron cooling of the energetic electrons. This is done by comparing the PP emission to the radio synchrotron emission with the implied minimal value of the magnetic field that is required to cool the electrons in times shorter than the SNR age $t$.

\subsection{Energy cutoffs}\label{sec:Cutoffs}
Consider a SNR with the following parameters:
Energy in shocked matter $E=10^{51} E_{51}\erg$, radius $R=10R_1\pc$, age $t=\tkyr\kyr$ shock velocity $\vv_s=1000\vv_8\km\se^{-1}$, ambient medium density of $n=n_{0}\cm^{-3}$ at a distance of $d=\dkpc\kpc$. The distance to the SNR is related to the radius by $\dkpc\approx R_1/\te^\circ$ where $\te^\circ$ is the angular diameter of the SNR on the sky in degrees.
The shock velocity, age and radius are related by $\vv_s=\al R/t\approx 10^9\al R_1\tkyr^{-1}\cm\se^{-1}$ with $0.4<\al<1$, the lower limit obtained for Sedov-Taylor (ST) expansion and the upper limit for free expansion (FE).

We assume that electrons and protons are accelerated to power law spectra [cf. Eq.~\eqref{eq:Powerlaw}] with $p\approx2$ up to cutoff energies, $\vep_{e,\text{max}}$ and $\vep_{p,\max}$ respectively, with a possible cooling break in the electron spectrum at $m_pc^2<\vep_{e,\text{break}}<\vep_{e,\text{max}}$ beyond which the power law index is $p+1$.

Both electron and proton energies are limited by the finite available acceleration time due to the finite SNR age. The maximal proton or electron energy due to the finite time satisfies:
\begin{equation}\label{eq:TimeLimit1}
t_{\text{acc}}(\vep_{\text{max}})=t,\end{equation}
where $t_{\text{acc}}(\vep)$ is the time it takes electrons or protons to reach energy $\vep$ (assumed to be equal for protons and electrons).
The maximal energy of accelerated electrons can also be limited by cooling, in which case we have
\begin{equation}\label{eq:CoolLimite1}
t_{\text{acc}}(\vep_{e,\text{max}})=t_{\text{cool}}(\vep_{e,\text{max}}),\end{equation}
where $t_{\text{cool}}(\vep_e)$ is the cooling time of electrons with energy $\vep_e$.
In the later case, a cooling break is expected at an energy $\vep_{e,\text{break}}$ satisfying
\begin{equation}\label{eq:CoolBreak1}
t_{\text{cool}}(\vep_{e,\text{break}})=t.\end{equation}

The acceleration time, $t_{\text{acc}}(\vep)$, can be approximated by,
\begin{equation}\label{eq:Tacc1}
t_{\text{acc}}\approx\frac{t_{\text{cycle}}}{\frac43\De\bt},\end{equation}
where $\De\bt=(\vs-u_d)/c$, $u_d\approx \vs/4$ is the downstream velocity and $t_{\text{cycle}}$ is the shock crossing cycle time.
Assuming the downstream residence time dominates the cycle time, we have
\begin{equation}\label{eq:Tcycle}
t_{\text{cycle}}=\frac{4D_d}{u_dc},\end{equation}
where $D_d$ is the downstream diffusion coefficient. The diffusion coefficient can be expressed as
\begin{equation}\label{eq:Diff}
D_d=\frac{\xi \vep c}{3qB},\end{equation}
where $\xi$ is a dimensionless coefficient that satisfies $\xi\geq1$ with $\xi=1$ obtained in the Bhom diffusion limit.
Using equations \eqref{eq:Tacc1}-\eqref{eq:Diff}, we can write the acceleration time as
\begin{equation}\label{eq:Tacc2}
t_{\text{acc}}\approx\frac{16}{3}\xi\frac{\vep}{qB\bt_s\vs}.\end{equation}

Using Eqs. \eqref{eq:TimeLimit1} and \eqref{eq:Tacc2}, the maximal energy due do to the limited SNR age can be expressed as
\begin{align}\label{eq:TimeLimit2}
\vep_{e,p}(t_{\text{acc}}=t)\approx \frac3{16} \xi^{-1}\al qB\bt_sR\approx 60~\xi^{-1}\al B_{-5}\vv_8R_1\TeV.
\end{align}

The electron cooling time due to synchrotron emission is given by
\begin{align}\label{eq:Tcool}
t_{\text{cool}}=\frac{\vep_e}{(4/3)\gamma_e^2\sig_TcU_B}\approx 10\left(\frac{\vep_e}{10\TeV}\right)^{-1}B_{-5}^{-2}\kyr
\end{align}
[we neglect the effect of IC which corresponds to a cooling time of $\approx 100(\vep_e/10\TeV)^{-1}\kyr$ which we assume is much larger than the SNR age].
We thus expect a cutoff to the synchrotron and IC spectra at photon energies given by
\begin{align}\label{eq:hnuCutoffTime}
&h\nu_{\Syn}(t_{\text{acc}}=t)\sim 1B_{-5}^3\vv_8^2R_1^2\xi^{-2}\al^2\keV,\cr
&h\nu_{\IC}(t_{\text{acc}}=t)\sim 10B_{-5}^2\vv_8^2R_1^2\xi^{-2}\al^2\TeV,
\end{align}
and a cutoff to the PP spectrum at photon energies given by
\begin{align}\label{eq:hnuCutoffTimePP}
&h\nu_{\PP}(t_{\text{acc}}=t)\sim 10B_{-5}\vv_8R_1\xi^{-1}\al\TeV.
\end{align}

Electron cooling will be relevant at the energy given by Eq. \eqref{eq:TimeLimit2} for strong enough magnetic fields,
\begin{align}\label{eq:MinBCool}
B>10\left(\frac{\xi}{\al^2R_1^2}\right)^{1/3}\muG.
\end{align}
If Eq.~\eqref{eq:MinBCool} is satisfied, there would be a spectral break at synchrotron and IC photon energies given by:
\begin{align}\label{eq:hnuBreak}
&h\nu_{\Syn}(t_{\text{cool}}=t)\approx 3B_{-5}^{-3}\tkyr^{-2}\keV,\cr
&h\nu_{\IC}(t_{\text{cool}}=t)\approx 40B_{-5}^{-4}\tkyr^{-2}\TeV,
\end{align}
 and a cutoff at photon energies of
\begin{align}\label{eq:hnuCutoffCool}
&h\nu_{\Syn}(t_{\text{cool}}=t_{\text{acc}})\approx\frac{3^3}{2^7}\xi^{-1}\al_e^{-1}m_e\vv_s^2\sim \xi^{-1}0.15\vv_8^2\keV,\cr
&h\nu_{\IC}(t_{\text{cool}}=t_{\text{acc}})\sim\xi^{-1}2\vv_8^2B_{-5}^{-1}\TeV.
\end{align}

The synchrotron and IC flux per logarithmic photon energy at photon energies above the break and below the cutoff would be suppressed by a factor of
\begin{equation}\label{eq:CoolingSupp}
\frac{\nu L_\nu~(\text{with cooling break})}{\nu L_\nu~(\text{no break})}\sim\left[\frac{h\nu}
{h\nu(t_{\text{cool}}=t)}\right]^{-1/2}\end{equation}
compared to the flux that would be emitted by a power-law without a break.

\subsection{SNR dynamics}\label{sec:Dynamics}
We now focus on the Sedov-Taylor (ST) phase. The SNR enters the ST phase when the mass of the swept up ambient medium,
\begin{align}\label{eq:SweptMass}
M_{\text{swept}}\sim 100R_1^3n_0 M_{\odot},
\end{align}
is larger than the ejecta's mass.
In this case we have:
\begin{align}\label{eq:STphase}
&\al=0.4,\cr
&t\approx \sqrt{\frac{0.5\rho R^5}{E}}\sim 5 R_1^{5/2}n_0^{1/2}E_{51}^{-1/2}\kyr,\cr
&\vv_{s}=0.4 R/t\sim 10^8E_{51}^{1/2}R_1^{-3/2}n_0^{-1/2}\cm\se^{-1}.
\end{align}

Substituting Eq.~\eqref{eq:STphase} in Eqs.~\eqref{eq:hnuCutoffTime}-\eqref{eq:hnuCutoffCool} we obtain:
\begin{align}\label{eq:hnuSynST}
&h\nu_{\Syn}(t_{\text{cool}}=t)_{\ST}\sim 0.15E_{51}B_{-5}^{-3}R_1^{-5}n_0^{-1}\keV,\cr
&h\nu_{\Syn}(t_{\text{cool}}=t_{\text{acc}})_{\ST}\sim 0.1\xi^{-1}E_{51}R_1^{-3}n_0^{-1}\keV,\cr
&h\nu_{\Syn}(t_{\text{acc}}=t)_{\ST}\sim 0.1 \xi^{-2}B_{-5}^3E_{51}R_1^{-1}n_0^{-1}\keV,
\end{align}
\begin{align}\label{eq:hnuICST}
&h\nu_{\IC}(t_{\text{cool}}=t)_{\ST}\sim 2B_{-5}^{-4}E_{51}R_1^{-5}n_0^{-1}\TeV,\cr
&h\nu_{\IC}(t_{\text{cool}}=t_{\text{acc}})_{\ST}\sim 1.5\xi^{-1}B_{-5}^{-1}E_{51}R_1^{-3}n_0^{-1}\TeV,\cr
&h\nu_{\IC}(t_{\text{acc}}=t)_{\ST}\sim 1\xi^{-2}B_{-5}^2E_{51}R_1^{-1}n_0^{-1}\TeV,
\end{align}
and
\begin{align}\label{eq:hnuPPST}
&h\nu_{\PP}(t_{\text{acc}}=t)_{\ST}\sim 3\xi^{-1}B_{-5}E_{51}^{1/2}R_1^{-1/2}n_0^{-1/2}\TeV.
\end{align}

\subsection{SNRs with observable non-thermal X-rays}\label{sec:NonThermal}
For SNRs with observable non-thermal synchrotron X-rays, we can find a lower limit to the shock velocity by demanding that there will be no cooling cutoff for photons with energies smaller than $h\nu_X=\nukeV\keV$, i.e.  $h\nu_{\Syn}(t_{\text{cool}}=t_{\text{acc}})>h\nu_X$. Using Eq.\eqref{eq:hnuCutoffCool}, this can be written as
\begin{align}\label{eq:MinVsNTX}
&\vv_s>3\X10^{8}\xi^{1/2} \nukeV^{1/2}\cm\se^{-1}.\end{align}
The minimal velocity constraint has several implications.
First, this can be used to obtain a minimal value for the proton temperature in the downstream. Comparing Eqs. \eqref{eq:hnuCutoffCool} and \eqref{eq:Tp}, we find:
\begin{equation}\label{eq:MinTpNTX}
T_p>\frac{8}{9}\xi\al_e\frac{m_p}{m_e}h\nu_X\sim 10\xi h\nu_X.\end{equation}
Second, assuming that the shock velocity is not much larger than $3000 \vv_{8.5}\km\se^{-1}$, the diffusion coefficient cannot be much larger than the Bohm limit ($\xi=1$),
\begin{align}
\xi\lesssim 1\vv_{8.5}^2\nukeV^{-1}.\end{align}
Under this assumption, the proton temperature is constrained by:
\begin{equation}\label{eq:TpNTX}
10\xi h\nu_{X}\lesssim T_p \lesssim 20\vv_{8.5}^2\keV.\end{equation}

Third, using $E\gtrsim 3\rho \vv_s^2 R^3$ (which is valid for both the ST and FE phases), we find a lower limit to the ambient medium density of
\begin{align}\label{eq:MaxnNTX}
n<0.1\frac{E_{51}}{R_1^3\xi}\nukeV^{-1}\cm^{-3}.
\end{align}

Next we compare the cutoff in the IC emission to the cutoff in the synchrotron radiation.
The energies of photons emitted by electrons through IC and Synchrotron are both proportional to the square of the Lorentz factor of the emitting electrons. The ratio of photon energies emitted through IC by electrons to the photon energies emitted through synchrotron by the same electrons is approximately given by:
\begin{align}\label{eq:ICSynXFreq}
&\frac{h\nu_{\IC}}{h\nu_{\Syn}}\approx 3T_{\text{CMB}}\frac{4\pi m_ec}{qB}\approx 10^{10}B_{-5}^{-1}.
\end{align}
The ratio of IC to synchrotron power, emitted by the same electrons, is approximately given by:
\begin{align}\label{eq:ICSynX}
&\frac{\nu_{\IC}L_{\nu_{\IC}}}{\nu_{\Syn}L_{\nu_{\Syn}}}\approx \frac{U_{\text{CMB}}}{U_B}\approx0.1B_{-5}^{-2}.
\end{align}

In particular, the photon energies where the IC and the synchrotron luminosities are cutoff should satisfy Eq. \eqref{eq:ICSynXFreq}, and the luminosity values at these photon energies should satisfy Eq. \eqref{eq:ICSynX} (this is true in principle for any feature in the spectrum).
We can use both equations to write a constraint that does not depend on the value of the magnetic field (or the acceleration mechanism):
\begin{align}\label{eq:ICSynCon}
&\frac{h\nu_{IC,\text{cutoff}}}{h\nu_{Syn,\text{cutoff}}}\sim 3\X10^{10}\sqrt{\frac{\nu_{IC}L_{\nu_{IC}}|_{\nu_{IC,\text{cutoff}}}}{\nu_{\Syn}L_{\nu_{\Syn}}|_{\nu_X,\text{cutoff}}}}.
\end{align}

A note of caution is in order regarding equation \eqref{eq:ICSynCon}. A 'cutoff' frequency is not a well defined quantity in general. For known functional forms, prescriptions for defining a specific frequency can be given. The precise value of the numerical coefficient in \eqref{eq:ICSynCon} may be somewhat different for different prescriptions. In addition it should be noted that while the IC spectrum of a single electron has a sharp cutoff (photons with energies larger than the initial electron energy cannot be generated), the synchrotron spectrum cuts-off exponentially, resulting in different photon spectra for given cutoff forms. Taking this into consideration and since the precise electron spectrum is not known, the cutoff frequencies are defined only to within an order of magnitude.

\subsection{Suppression of IC due to radiative cooling}\label{sec:ICSupp}
The calculated expected ratio given by Eq.~\eqref{eq:ICPP} is valid as long as there is no significant suppression of the electron population due to cooling.
Electrons responsible for $\TeV$ IC emission have Lorentz factors of approximately ${\gamma\sim(\TeV\nuTeV/3T_{\text{CMB}})^{1/2}\sim 4\X10^7\nuTeV^{1/2}}$ and a corresponding cooling time of [cf. Eq.\eqref{eq:Tcool}]:
\begin{align}
t_{\text{cool}}=6 \nuTeV^{-1/2}B_{-5}^{-2}\kyr.
\end{align}

Cooling will affect these electrons only if the cooling time is shorter than the lifetime $t$ of the SNR which would be true only if the typical magnetic field is large enough:
\begin{align}\label{eq:MinBICSupp}
B\gtrsim 30 \nuTeV^{-1/4}\tkyr^{-1/2}\muG.
\end{align}

A larger magnetic field would imply stronger synchrotron emission. We can use this to write a constraint on the $\TeV$ PP emission in case the IC emission is suppressed.
Assuming that the electrons responsible for the IC $\TeV$ emission were suppressed by cooling, we can use Eqs.~\eqref{eq:MinBICSupp} and \eqref{eq:PPSynGHz} to obtain:
\begin{align}\label{eq:PPSynICSupp}
&\frac{\nu_{\gamma} L_{\nu_{\gamma}~\PP}}{\nu_{R}L_{\nu_{R}~\Syn}}< 10K_{ep,-2}^{-1}\tkyr^{3/4}n_0\nuTeV^{3/8}.
\end{align}

Another constraint can be derived by comparing the PP emission to the Synchrotron radiation at X-ray frequencies assuming that the electrons emitting the X-rays are also affected by cooling. This assumption is reasonable since we assume that electrons responsible for $\TeV$ IC emission are affected by cooling, and these electrons are responsible for synchrotron radiation of photons with energies $h\nu\gtrsim100B_{-5}\eV$ [cf. Eq. \eqref{eq:ICSynXFreq}].
Using Eqs. \eqref{eq:PPSynGHz} and \eqref{eq:CoolingSupp}, the ratio of the PP flux to the X-ray flux in the frequency range between the cooling break given by Eq. \eqref{eq:hnuBreak} and the cooling cutoff given by Eq. \eqref{eq:hnuCutoffCool} (for a spectrum with $p\geq2$ this is the maximum value of $\nu L\nu _{\Syn}$) is given by
\begin{align}\label{eq:PPSynXICSupp}
&\frac{\nu_{\gamma} L_{\nu_{\gamma}~\PP}}{\nu_{X}L_{\nu_{X}~\Syn,\max}}\sim 0.4 n~\sig_{pp}^{\text{inel}}c~tK_{ep}^{-1}\sim 1.5\X10^{-3}K_{ep,-2}\tkyr n_0,\cr
\end{align}
where we assumed $p=2$. This equation has a weak dependence on $p$ since the X-ray emitting electrons have energies that are similar to the $\TeV$ $\gamma$-ray emitting protons.
Equation \eqref{eq:PPSynXICSupp} has the following simple interpretation.
Suppose that the amount of protons per unit energy and unit time that are being accelerated by the shock is given by $Q(\vep)$. The amount of protons per unit energy at an age $t$ is roughly $dN/d\vep\sim Q(\vep)t$ and so the PP luminosity per logarithmic frequency is roughly given by [cf. Eq. \eqref{eq:PPExact}] $\nu L\nu_{\PP}\sim 0.2 Q(\vep)\vep^2t n c \sig^{\text{inel}}_{pp}$. The electron injection rate at electron energies of $\vep$ is $K_{ep}Q(\vep)$. As the electrons are constantly being cooled, the energy input in accelerated electrons is equal to the energy emitted in synchrotron radiation. The X-ray synchrotron luminosity per logarithmic frequency is thus roughly $\nu L\nu_{\Syn~\text{cooled}}\sim 0.5 K_{ep}Q(\vep)\vep^2$ (the factor of 0.5 comes from the fact that the logarithmic interval in photon energies is twice that of the emitting electrons due to the $\nu\propto \gamma^2$ dependence). The ratio of these expressions is equal to the result in Eq. \eqref{eq:PPSynXICSupp}.

\section{Lower limit on $K_{ep}$ from extragalactic SNRs}\label{sec:M33}
In this section we find a lower limit for $K_{ep}$ using the observed radio fluxes from large SNRs in M33
assuming that the value of $K_{ep}$ does not vary significantly between SNRs.

One way to to estimate the amount of accelerated electrons is through the radio synchrotron emission. The radio luminosity is determined by the energy in accelerated electrons and by the magnetic field value. The amount of energy in accelerated electrons cannot be deduced if the value of the magnetic field is not known.
An upper limit to the magnetic field is given by the requirement that the magnetic field does not exceed equipartition.
Here we assume that the fraction $\ep_B$ of the thermal energy carried by the magnetic field behind the shock does not significantly exceed $\ep_B\sim 0.1$. In addition we assume that the fraction $\eta_p\sim\ep_p/2$ of the total energy carried by relativistic protons does not significantly exceed $\eta_p\sim0.1$.
Using equations \eqref{eq:Powerlaw} and \eqref{eq:SynExact} we can approximate the expected luminosity at $1\GHz$ by:
\begin{align}\label{eq:SynLum}
&L_{\nu,\Syn}(\GHz)\approx 4\X10^{22}K_{ep,-2}\eta_{p,-1}E_{51}B_{-5}^{3/2}\cr
&\X\Lm_{p,1}^{-1}(5B_{-5}^{-1/2})^{-(p-2)}\erg\cm^{-2}\se^{-1}\Hz^{-1}
\end{align}
where $\eta_p=0.1\eta_{p,-1}$. For an assumed maximal Lorentz factor $\gamma_{p,\text{max}}\sim 10^5$, the factor in the second line of Eq.~\eqref{eq:SynLum} equals $\approx 0.9$ for $p=2$ and $\approx 1.6$ for $p=2.2$, and will be ignored henceforth.

The ratio of the magnetic to thermal energies behind the shock in the Sedov-Taylor phase can be approximated by:
\begin{align}\label{eq:epB}
\ep_B\approx\frac{B^2/8\pi}{\rho\vv_s^2}\approx 3\frac{B^2}{8\pi}R^3E^{-1}\approx 4\X10^{-4}B_{-5}^2R_1^3E_{51}^{-1}.
\end{align}
Extracting the magnetic field from Eq. \eqref{eq:epB} and substituting it in Eq. \eqref{eq:SynLum} we have

\begin{align}\label{eq:SynLumepB}
&L_{\nu,\Syn}(\GHz)\approx 3\X10^{24}K_{ep,-2}\eta_{p,-1}E_{51}^{7/4}\ep_{B,-1}^{3/4}R_1^{-9/4} \erg\se^{-1}\Hz^{-1}.
\end{align}

Radio luminosities of SNRs with known distances in nearby galaxies (including the milky way) are summarized by \citet{{Arbutina04}}, and virtually all have luminosities greatly exceeding $3\X10^{22}R_1^{-9/4} \erg\se^{-1}\Hz^{-1}$, the typical value expected from Eq. \eqref{eq:SynLumepB} for $K_{ep}=10^{-4}$. However, we should stress that it is dangerous to reach conclusions based on such comparisons, since the observed luminosities are limited from below by the detectors' sensativities.
Here we focus on a sample of SNRs in M33 which is perhaps the most complete sample of radio SNRs with known distances in a single galaxy \citep{Gordon99}.

Using Eq. \eqref{eq:SweptMass} we see that SNRs with radii larger than
\begin{align}
R\gtrsim 2 \left(\frac{M_{\text{ej}}}{10M_{\odot}}\right)^{1/3}n_0^{-1/3}\pc
\end{align}
are in the ST phase. The smallest SNR in the sample has a radius of $R\sim 5\pc$ and most SNRs in the sample have radii $R>10\pc$. It is thus reasonable to assume that the SNRs in the sample are in the ST expansion phase. In fact, \citet{Gordon98} have shown that the radii distribution function of a larger optical SNR sample that includes the radio SNR sample is consistent with ST expansion (and is inconsistent with free expansion).

\begin{figure}[h]
\epsscale{1} \plotone{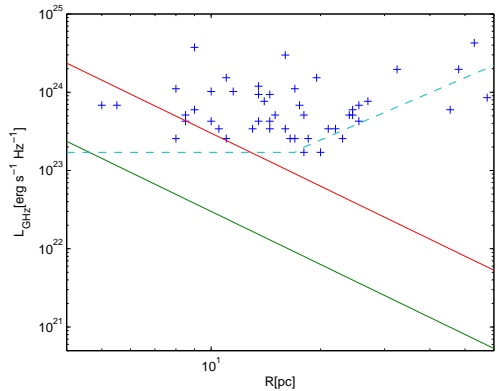}
\caption{\label{fig:M33} Radio 20cm luminosity of SNRs in M33 \citep{Gordon99}. The dashed line is the observation flux (total and density) limit. The full lines are given by Eq. \eqref{eq:SynLumepB} with $K_{ep}=10^{-4}$ (lower, green) and $K_{ep}=10^{-3}$ (higher, red).}
\end{figure}

The luminosities of the observed SNRs in M33 are shown in figure \ref{fig:M33} along with the observational threshold (dashed line) and the expected limits according to Eq. \eqref{eq:SynLumepB} corresponding to $K_{ep}=10^{-4}$ (lower, green) and $K_{ep}=10^{-3}$ (higher, red), adopting a distance of $d=840\kpc$ to M33.

As stressed by \citet{Gordon99}, there are probably unobserved SNRs with luminosities that fall beneath the observational threshold. In fact, there are about twice as many SNRs seen in optical wavelengths \citep[][the factor being roughly radius independent, e.g. 8, 23 and 34 SNRs in the radio sample with radii $R<10,15,20\pc$  respectively compared to 15,42 and 67 SNRs respectively in the optical sample]{Gordon98}.
Assuming the optical sample is not far from completeness, it is reasonable that roughly half of the SNRs are missed in the radio sample \citep[this is true for $R\lesssim20\pc$, while for $R\gtrsim20\pc$ the optical sample is probably incomplete,][]{Gordon98}.
Still, it is quite clear from figure \ref{fig:M33} that the luminosity implied by a value $K_{ep}=10^{-4}$ is lower than the typical luminosity of large remnants by at least an order of magnitude.
As an illustration, in order to reconcile a value of $K_{ep}\sim10^{-4}$ with the four SNRs observed with radius $R\approx 50\pc$ and luminosity $L_{\nu}(\GHz)\sim 10^{24}\erg\se^{-1}\Hz^{-1}$, their energies would have to be unreasonably high:
\begin{align}
E\sim 5\X10^{52}\left(\frac{R}{50\pc}\right)^{9/7}K_{ep,-4}^{-4/7}\eta_{p,-1}^{-4/7}\ep_{B,-1}^{-3/7}\cr
\X\left(\frac{L_{\nu}(\GHz)}{10^{24}\erg\se^{-1}\Hz^{-1}}\right)^{4/7}\erg.
\end{align}
Note that SNe with energies that are larger than $10^{52}\erg$ (termed Hypernovae) have been detected \citep[see e.g.][and references therin]{Nomoto06}. However, the estimated fraction of core-collapse SNe that belong to this group is of order $10^{-3}$ \citep{Podsiadlowski04} and thus having 4 SNRs with energies exceeding $10^{52}\erg$ among the $\sim 100$ SNRs in M33 is unlikely.

Using Eq.~\eqref{eq:SynLumepB}, we conclude that $K_{ep}\gtrsim 10^{-3}$ is a reasonable lower limit and that $K_{ep}\sim10^{-4}$ can be conservatively ruled out.

A possible caveat in the arguments in this section comes from the fact that it is possible that the ambient CR electrons that have been swept up by the shock have a considerable contribution to the Synchrotron emission \citep{Anderson93}. The arguments in this section will nevertheless remain valid in this case too, provided the ratio of the accelerated electrons and protons populations, including the CR contributions, are similar for different SNRs.  We note that assuming that the cosmic rays that enter the shock are reaccelerated by DSA, the shape of the spectrum of the population of relativistic particles will approach a power law and will not be affected by the distribution of the CRs in the ISM \citep{Drury94}.

\section{Application to \SNRa{} and \SNRb{}}\label{sec:SNRSab}
Here we apply the results of sections \sref{sec:OneZone}-\sref{sec:M33} to show that the broad-band spectrum of the SNRs, \SNRa{} and \SNRb{} is inconsistent with a PP origin and is consistent with an IC origin of the $\gtrsim$ TeV emission.
First we summarize in \sref{sec:SNRsabChar} the broadband observations of these SNRs.
Next, we show in \sref{sec:SNRsabPP} that a PP source for the observed $\gamma$-ray flux is inconsistent with the broad-band emission in these SNRs.
We then show in \sref{sec:SNRsabIC} that an IC source for the observed $\gamma$-ray flux is consistent with all observations. We show that the contribution of the PP $\gamma$-ray emission is negligible and argue that the neutrino emission from these SNRs is probably too low to be detected by current and planned neutrino telescopes.
Finally, we compare in \sref{sec:Comparison2} the results presented here to previous studies. In particular we discuss previous claims against an IC source of the $\gamma$-rays.

\subsection{Characteristics of \SNRa{} and \SNRb{}}\label{sec:SNRsabChar}
The observations of these SNRs are described by \citet{Aharonian06,Aharonian07} and references therein. Some of the main features are summarized below. In many ways these two shell type SNRs are similar . Both have comparable radio and $\TeV$ fluxes,
\begin{align}
&\nu f_{\nu}|_{\text{GHz}}\approx \text{few}~ 10^{-13}\erg\cm^{-2}\se^{-1},\cr &\nu f_{\nu}|_{\text{TeV}}\approx \text{few}~
10^{-11} \erg\cm^{-2}\se^{-1},
\end{align}
span similar angles on the sky ($\te\approx1,2^\circ$ respectively), have non thermal X-ray emission, which is consistent with a cutoff frequency of order $h\nu_{\text{cutoff}}\lesssim \keV$.
The gamma ray energy flux is consistent in both SNRs with a power law $\nu f_{\nu}\propto \nu^{0}$ and an exponential cutoff at photon energies of $\sim 10 \TeV$ \citep[for \SNRb{} the detection of the cutoff is less clear, ][]{Aharonian07}.

Perhaps the main difference is in the $h\nu\sim 1\keV$ X-ray flux which is larger for \SNRa{} by a factor of about 5,
$\nu f_{\nu}|_{\keV}\approx \text{few}~10^{-10},\text{few}~10^{-11}\erg\cm^{-2}\se^{-1}$ respectively.

One of the main characteristics of these SNRs is a non thermal dominated X-ray emission. The lack of observable thermal radiation can be used to obtain an upper bound on the value of the ambient density. Number densities considerably smaller than $1\cm^{-3}$ where obtained \citep{Slane99,Slane01,Pannuti03,Cassam04}, which in turn constrain the amount of proton-proton collisions.
For \SNRa{}, limits on $n$ from lack of thermal radiation of $n<0.3\cm^{-3}(\dkpc/6)^{-1/2}$, $n\approx0.05-0.07\cm^{-3}(\dkpc/6)^{-1/2}$ and $n<0.02\dkpc^{-1/2}\cm^{-3}$ were obtained by \citet{Slane99,Pannuti03} and \citet{Cassam04} respectively.
For \SNRb{} a limit on $n$ from lack of thermal radiation (for temperature greater than $1\keV$) of $n<0.03 \dkpc^{-1/2}\cm^{-3}$ was obtained by \citet{Slane01}.

There have been claims that \SNRa{} is interacting with molecular clouds \citep[][at ~6kpc and ~1kpc respectively]{Slane99,Fukui03}.  Interaction with molecular clouds of both SNRs is unlikely given the low densities implied from lack of thermal radiation and the observed roughly homogenous emission \citep{Aharonian06,Aharonian07}. The positive TeV to CO line emission correlation that was claimed for \SNRa{} is not convincing since the CO intensity changes by some two orders of magnitude while the TeV changes by a factor of ~2 \citep[average to peak,][]{Aharonian06,Aharonian07}.  In any case, interaction with molecular clouds cannot account for the entire emission and we will ignore this possibility henceforth.

Distance and age estimates for these remnants are inconclusive \citep[][and references within]{Aharonian06,Aharonian07}.
We think that it is worth mentioning that claims that the distance to these SNRs is $d\lesssim 1\kpc$ \citep[e.g.][]{Fukui03,Aschenbach99} require some coincidence since the galactic latitude of both SNRs is $b\lesssim 1^\circ$, ($b=0.5,1.2$ for \SNRa{} and \SNRb{} respectively)  whereas the SNRs at this distance should be distributed in the range $\abs{b}\lesssim 10^\circ$ assuming SNRs are distributed homogenously throughout the galactic gaseous disk hight. For \SNRa{} the coincidence that is required is more extreme since this SNR lies in the direction of the galactic center, $b=0.5^\circ,l=347^\circ$, close to a 'hole' in the galactic CO line emission \citep{Slane99,Moriguchi05}. These positions on the sky may not be coincidental if these SNRs are farther away- a few kpcs from us (as most SNRs are). On the other hand, we note that such a coincidence is certainly possible and we do not assume in what follows that the distance to these remnants is larger than $1\kpc$.

\subsection{Upper bounds on PP emission}\label{sec:SNRsabPP}
We first consider the constraints on the PP emission resulting from the comparison of the PP emission to the IC and Synchrotron non-thermal emission.
By inserting $K_{ep}\sim 10^{-2}$ and $n\lesssim 0.1\cm^{-3}$ in Eq. \eqref{eq:ICPP} we see that unless the IC emission is suppressed by cooling, the $\gtrsim 1\TeV$ emission in these SNRs is completely dominated by IC.
Irrespective of cooling, the ratio of PP $\gtrsim1\TeV$ emission to the synchrotron radio emission is given by Eq. \eqref{eq:PPSynGHz}.
A lower limit for the magnetic field is given by \eqref{eq:MinBICSupp} for the case where IC emission is suppressed by cooling.
Alternatively, a lower limit of $B\gtrsim 10\muG$ can be derived by demanding that the IC emission generated by the electrons that emit the observed X-ray synchrotron emission does not exceed the observed gamma ray emission [using Eqs. \eqref{eq:ICSynXFreq} and \eqref{eq:ICSynX}, see e.g. \citet{Aharonian06}].
By inserting $n=0.1n_{-1}\cm^{-3}$ and $B\gtrsim 10B_{-5}\muG$ in Eqs. \eqref{eq:PPSynGHz} and \eqref{eq:PPSynICSupp} and assuming $p\geq2$, we find that
\begin{align}\label{eq:PPSynGHzSNRsab}
\frac{\nu_{\gamma} L_{\nu_{\gamma}~\PP}}{\nuGHz L_{\nuGHz~\Syn}}\lesssim 5K_{ep,-2}^{-1}B_{-5}^{-3/2}n_{-1}
\end{align}
and
\begin{align}\label{eq:PPSynICSuppSNRsab}
\frac{\nu_{\gamma} L_{\nu_{\gamma}~\PP}}{\nuGHz L_{\nuGHz~\Syn}}\lesssim 1K_{ep,-2}^{-1}\tkyr^{3/4}n_{-1}\nuTeV^{3/8},
\end{align}
with the later equation applicable if the IC emission of photons with energy $h\nu=\nuTeV\TeV$ is suppressed by cooling.
Comparing this to the observed ratio of fluxes per logarithmic frequency at $h\nu=1\TeV$ and $\nu=1\GHz$, which for \SNRa{} and \SNRb{} is $\nu f_\nu(\TeV)/\nu f_{\nu}(\GHz)\sim 100$, we see that the contribution of the PP $\TeV$ emission is negligible compared to the total $\gtrsim1\TeV$ emission.

In case that the synchrotron X-rays are also affected by cooling (this is likely if the IC emission is suppressed by cooling), by using equation \eqref{eq:PPSynXICSupp} we find that
\begin{align}\label{eq:PPSynXICSuppSNRsab}
&\frac{\nu_{\gamma} L_{\nu_{\gamma}~\PP}}{\nu_{X}L_{\nu_{X}~\Syn,\max}}\sim 1.5\X10^{-4}K_{ep,-2}\tkyr n_{-1}.\cr
\end{align}
Comparing this to the observed ratio of fluxes per logarithmic frequency at $h\nu=1\TeV$ and $h\nu\sim 1\keV$, which for \SNRa{} and \SNRb{} are $\nu f_\nu(\TeV)/\nu f_{\nu}(\keV)\sim 10$ and $2$ respectively, and assuming that the maximum Synchrotron luminosity cannot be much higher at lower frequencies, we see again that the contribution of the PP $\TeV$ emission is negligible compared to the total $\gtrsim1\TeV$ emission.

We have assumed above that $K_{ep}\sim 10^{-2}$, in accordance with the local ratio of CR electrons to protons and with section \sref{sec:M33}. It should be emphasized that a direct estimation of the value of $K_{ep}$ using the arguments of section \sref{sec:M33} is not possible for these SNRs, since their radio luminosity is not known (due to the uncertain distances) and since the value of $\ep_B$ is not known for these remnants (the expected value of $\ep_B$ is discussed in \sref{sec:Discussion}). In \sref{sec:SNRsabIC} it is shown
using Eq. \eqref{eq:SynFlux} that a magnetic field value of $B\sim 10\muG$, required in the IC scenario, and a value for $K_{ep}$ of $K_{ep}\sim10^{-2}$ are consistent with the observed radio flux for a distance of $\sim1\kpc$.

We next consider the constraint on the PP emission resulting from the comparison of the PP emission to the TB X-ray emission.
The ratio of $\gtrsim1\TeV$ PP Luminosity to TB Luminosity is given by Eq. \eqref{eq:PPTB2}.

Constraints on the shock velocity and more importantly the post-shock proton temperature for SNRs with observable non-thermal X-ray radiation are given by Eqs. \eqref{eq:MinVsNTX} and \eqref{eq:TpNTX}
\begin{align}\label{eq:MinVsNTXSNRsab}
&\vv_s>3\X10^{8}\xi^{1/2} \nukeV^{1/2}\cm\se^{-1},\end{align}
and
\begin{equation}\label{eq:TpNTXSNRsab}
10\xi\nukeV\lesssim T_p \lesssim 20\vv_{8.5}^2\keV\end{equation}
respectively (the upper limit to $T_p$ results from the assumption $\vs\lesssim 3000\vv_{8.5}^2\km\se^{-1}$), where we assumed that the cutoff in the X-ray spectrum is at $h\nu_{\text{cutoff}}=\nukeV\keV$. $\xi$ is the inverse of the ratio of the diffusion coefficient to the maximal alowable, Bhom-diffusion coefficient and is always larger than 1.
The same arguments led \citet{Berezhko06} to the conclusion that $\vv_s> 1.5\X10^{8}\cm\se^{-1}$ for \SNRa{} (the value they obtained from the broad-band fit is $\vv_s\approx1.8\X10^8\cm\se^{-1}$).

Substituting $T_p=10T_{p,1}\keV$ [following Eq. \eqref{eq:TpNTXSNRsab}] in Eq. \eqref{eq:PPTB2} and assuming $p\geq2$ we find:
\begin{align}\label{eq:PPTB2SNRsab}
\frac{\nu_{\gamma} L_{\nu_{\gamma}~\PP}}{\nu_{X}L_{\nu_{X}~\TB}|_{h\nu=\zt_eT_p}}\lesssim 10^{-2}\ep_{p,-1}\zt_{e}^{-1/2}T_{p,1}^{1/2}.
\end{align}
$\zt_e$ is the ratio of post-shock electron and proton temperatures and $\ep_{p}=0.1\ep_{p,-1}$ is the fraction of the thermal energy in accelerated protons.
Comparing Eq. \eqref{eq:PPTB2SNRsab} to the observed ratio of fluxes per logarithmic frequency at $h\nu=1\TeV$ and $h\nu=1\keV$, which for \SNRa{} and \SNRb{} is $\nu f_\nu(\TeV)/\nu f_{\nu}(\keV)\sim 10$ and $2$ respectively, we see that a PP origin of the $\gtrsim1\TeV$ is unlikely for \SNRa{} and not possible for \SNRb{} (since a TB flux greatly exceeding the observed X-ray flux would be implied) as long as there is significant collisionless electron heating $\zt_e\sim1$.

If there is no collisionless electron heating, we can use Eq. \eqref{eq:PPTBCoulomb} (with $T_p=10T_{p,1}\keV$ and $n=0.1n_{-1}\cm^{-3}$)
\begin{align}\label{eq:PPTBCoulomb2}
&\frac{\nu_{\gamma} L_{\nu_{\gamma}~\PP}}{\nu_{X}L_{\nu_{X}~\TB}|_{h\nu=T_e}}\lesssim
0.04
\ep_{p,-1}T_{p,1}^{4/5} \left(\lm_{ep,1.5}n_{-1}t_{\text{kyr}}
\right)^{-1/5}.
\end{align}
The electron temperature will be [see Eq. \eqref{eq:TeTpCoulomb}]:
\begin{equation}
T_e\gtrsim 0.6(\lm_{ep,1.5}n_{-1}T_{p,1})^{2/5} \keV.
\end{equation}
For a temperature of $T_e\gtrsim0.6\keV$ the thermal emission implied from Eq. \eqref{eq:PPTBCoulomb2} for a PP model, would likely be detectable in \SNRa{}, especially if we take into account that there would be line emissions that would have higher luminosities. In \SNRb{}, emission at frequencies below $1\keV$ might be hard to detect due to the high background of thermal radiation coming from the Vela SNR \citep{Slane01}.
We note that if the proton acceleration is very efficient $\ep_p\sim1$, there is no collisionless heating and the TB emission is not considerably lower than the observed non-thermal X-rays, a PP origin cannot be ruled out based on this argument alone for either SNR.
We conclude that the $\gtrsim 1\TeV$ photons from \SNRa{} and \SNRb{} are unlikely to be emitted by PP interactions and thus are likely emitted by IC scattering.

\subsection{IC scenario}\label{sec:SNRsabIC}
We next ask whether the broad-band spectrum of these SNRs is consistent with an IC source of the $\gamma$-rays.

First note that for both SNRs, the inferred cutoff in the synchrotron at $\sim 1\keV$ is consistent with the cutoff observed in the $\sim 10\TeV$ emission (we should note that for \SNRb{} there is only a sign of a cutoff, the uncertanties do not alow a firm conclusion) if we assume a magnetic field of order $10\muG$ \citep[][somewhat less for \SNRb{}]{Aharonian06,Aharonian07,Porter06}.
In particular Eq. \eqref{eq:ICSynCon} is satisfied (up to the uncertainties in the cutoff frequencies). This by itself can be considered as an indication of an IC source.

As the $\gamma$-ray observations extend somewhat below the cutoff, down to $\approx 0.3\TeV$, it is reasonable to compare the gamma ray emission directly with the radio emission, ignoring the possible suppression of the gamma ray flux due to cooling.
Comparing equation \eqref{eq:ICSynGHz} with the observed ratio of $\gamma$-ray to radio flux, $\nu f_\nu(\TeV)/\nu f_{\nu}(\GHz)\sim 100$, we see that the expected ratio (for $p=2$) is $5-10$ times larger than observed in these SNRs (larger values corresponding to \SNRb{}). This apparent discrepancy can be due to cooling suppression of the IC flux or due to a value of $p$ slightly larger than 2 (e.g. $p=2.2$ would result in a factor of 5) consistent with the assumptions made here (a lower observed ratio would be inconsistent).

We next note that for $n\sim 0.1\cm^{-3},R\sim10\pc$ and $E\sim10^{51}$ the expected cutoffs in the radio and $\gamma$-ray spectrum, Eq.~\eqref{eq:hnuSynST}, \eqref{eq:hnuICST} are consistent with the observed cutoffs and cooling may or may not be important.
The expected radio flux according to \eqref{eq:SynFlux} is consistent with the observed $\sim 1\GHz$ flux for the corresponding distance $d\sim1\kpc$.
We would like to emphasize that there are more free parameters than constraints and these values are not the only ones allowable by these constraints.

We conclude that the PP contribution to the $\gtrsim1\TeV$ flux is negligible and that an IC source for the $\gtrsim1\TeV$ flux is consistent with the observed broad-band spectrum.

Using Eqs. \eqref{eq:PPSynGHzSNRsab} and \eqref{eq:PPTB2SNRsab}, we see that the expected neutrino flux (being roughly equal to the $\gamma$-ray flux) is constrained for these SNRs to values
\begin{align}\label{eq:NeutrinoFlux}
\vep_{\nu}f_{\vep_{\nu}}\lesssim10^{-12}\erg\cm^{-2}\se^{-1}.
\end{align}
The neutrino detection rate per logarithmic neutrino energy by a neutrino detector with an area $A=A_{\text{km}^2}\km^{2}$ is given by
\begin{align}
&\vep_{\nu}\frac{d\dot N_{\nu}}{d\vep_{\nu}}=f_{\vep_{\nu}}P_{\nu\mu,~\text{water}}A\sim \cr &\sim0.2 \frac{\vep_{\nu}f_{\vep_{\nu}}}{10^{-12}\erg\se^{-1}\cm^{-2}}A_{\text{km}^2}\yr^{-1},\cr
\end{align}
where $P_{\nu\mu,~\text{water}}$ is the probability that a neutrino will interact with the water and produce a muon within a distance from the detector that is smaller than the muon cooling distance, and is approximately given by:
$P_{\nu\mu,~\text{water}}\sim10^{-6}\vep_{\nu, \text{TeV}}^{1}$. This flux is probably too low to be detected by current and planned neutrino observatories.

\subsection{Comparison with previous studies}\label{sec:Comparison2}
Next we compare the results presented in \sref{sec:SNRsabPP} and \sref{sec:SNRsabIC} to previous studies of these SNRs.

In \sref{sec:Comparison1} it was shown that Eqs. \eqref{eq:ICPP} and \eqref{eq:PPSynGHz} agree with the results of studies of \SNRa{} \citep{Aharonian06,Berezhko06,Porter06,Moraitis06} and \SNRb{} \citep{Enomoto06,Aharonian07} to within a factor of $\sim2$.
We note that all models in which the $\gamma$-ray emission is dominated by PP, avoided the implied excessive synchrotron emission (but not the implied excessive thermal X-ray Bremsstrahlung emission, see \sref{sec:SNRsabPP}) by assuming an extremely low value of $n^{-1}K_{ep}$, of $n^{-1}K_{ep}\lesssim 10^{-3}$. Such low values of $n^{-1}K_{ep}$ are not plausible since a high density $n\gg0.1$ is inconsistent with the lack of observed thermal X-ray emission and a low value of $K_{ep}\lesssim10^{-4}$ is inconsistent as shown in \sref{sec:M33}.

\subsubsection{Claims against IC for \SNRa{} and \SNRb{}}\label{sec:ClaimsNotIC}
We next discuss the main claims that were raised against an IC source for the gamma ray emission in $\SNRa{}$ and
$\SNRb{}$.

\textit{Low magnetic field:} As discussed in \sref{sec:SNRsabIC}, a magnetic field of $B\sim 10\muG$ is implied if the
gamma ray emission is due to IC. The value of the magnetic field was estimated to be much higher, of order $100\muG$ \citep{Berezhko06,Volk05,Bamba05b} by interpreting thin filaments observed in the X-ray images as the result of small cooling lengths of the emitting electrons. If true this would rule out IC as the source of the gamma ray emission. The thin filaments could alternatively be interpret as thin regions of enhanced magnetic field \citep[e.g. ][]{Pohl05} in which case the magnetic field cannot be estimated directly.
One way to distinguish between the interpretations is by comparing high resolution radio and X-ray images. The interpretation that the filaments are due to cooling of multi-$\TeV$ electrons implies that similar features should not be seen in the radio image since the electrons responsible for the radio emission hardly suffer from radiative cooling \citep{Vink03}. An X-ray to radio comparison was done in \cite{Lazendic04}, however the low resolution radio images do not allow a decisive conclusion.
We should note that the same arguments were used to deduce high magnetic fields in Tycho's SNR and the remnant of SN1006  \citep[for which a high resolution radio image exists, ][]{Dickel91,Rothenflug04}, while some of the thin filaments in the X-ray emission are clearly seen also in the radio images , [compare \citep{Dickel91} figure 1 to \citep{Bamba05} figure 1, and \citep{Rothenflug04} figure 1 to \citep{Bamba03} figure 1,  see also \citep{Cassam07}], a fact that was ignored by \citet{Bamba05} and \cite{Volk05}.

\textit{Detailed spectral shape:}
\citet{Berezhko06} claim that the observed X-ray flux cannot be properly fitted for a magnetic field of the order $\sim 10\muG$. \citet{Aharonian06} claim that the shape of the gamma ray spectrum in \SNRa{} does not coincide with IC since an electron spectrum chosen to fit the radio and X-ray observations produces a narrow peak in $\nu f_\nu$ in disagreement with the flat gamma ray spectrum observed. We do not see these claims as an inconsistency as the physics of the cutoff in the particle spectrum is not really known. For example, the assumed diffusion coefficient value is not known for all energies. If the magnetic field disturbances are generated by the accelerated particles, the spectrum at scales relevant to the particles with energies close to the cutoff scale may be different than for intermediate scales.
In addition, if the high energy end of the electron energy distribution is affected by synchrotron energy losses, a flat $\nu f_\nu\propto \nu^{0}$ IC spectrum would be expected.

We note that if the synchrotron peak was resolved, a more trustable comparison of IC and synchrotron spectrums could have been done as long as the effect of the interstellar infra-red radiation is negligible \citep[see ][]{Porter06}.

We also note that there is some inconsistency in the model parameters assumed by \citet{Aharonian06}. They assume an age of $1000\yr$, a distance of $1\kpc$ and an ambient density of $n=1\cm^{-3}$. For such a distance and density, the swept up mass is $M\approx (4\pi/3)R^3nm_p\sim 100 M_\odot$ which is clearly in the ST regime and implies an energy in the swept up material of $E\approx 0.5nm_pR^5t^{-2}\approx 10^{52}\erg$, which is rather large. Demanding an energy of $10^{51}\erg$, for example, would imply an age of about $3000\yr$, for which cooling in a magnetic field of $10\muG$ may be important (the effect of cooling would be to flatten the IC and synchrotron peaks).

\section{Discussion}\label{sec:Discussion}
In this paper we derived simple analytic tools for analyzing the radio, X-ray and $\gtrsim1\TeV$ $\gamma$-ray continuum emission mechanisms in shell-type SNRs. The emission mechanisms considered were Synchrotron, IC of CMB photons by accelerated electrons, proton-proton collisions of accelerated protons with ambient protons and thermal-Bremsstrahlung.
In \sref{sec:OneZone} we wrote down the luminosity ratios of these emission mechanisms (ignoring the energy cutoffs), Eqs. \eqref{eq:ICPP}, \eqref{eq:PPTB2}, \eqref{eq:PPSynGHz} and \eqref{eq:ICSynGHz}. These ratios are independent of the SNR energy and of the distance to the SNR.
In \sref{sec:EnergyCutoffs} we wrote down the (energy and distance dependent) expected cutoffs in the non-thermal radiation spectra, Eqs. \eqref{eq:hnuSynST}-\eqref{eq:hnuPPST}, due to cooling and limited SNR age assuming DSA as the acceleration mechanism and Sedov-Taylor evolution.
In addition we obtained an energy and distance independent constraint, Eq. \eqref{eq:PPSynICSupp}, for the PP flux in case the IC spectrum is suppressed, and an energy and distance independent lower limit for the proton temperature $T_p$ for SNRs in which non-thermal X-rays are observed, Eq.\eqref{eq:TpNTX}.
We note that the synchrotron cutoff due to cooling given by Eq. \eqref{eq:hnuCutoffCool} \citep[a similar expression was derived by][]{Berezhko04} naturally explains the fact that synchrotron emission does not extend to photon energies greatly exceeding $\sim\keV$ in known SNRs \citep[see e.g.][]{Reynolds99,Hendrick01}. This is simply because the shock velocities in SNRs do not greatly exceed a few thousand $\km \se^{-1}$.

In \sref{sec:M33} we derived a lower limit to the value of $K_{ep}$, the ratio of the number of accelerated electrons and the number of accelerated protons at a given energy, $K_{ep}>10^{-3}$, by studying the radio observations of SNRs in M33.
Here we assumed that the value of $K_{ep}$ (including the possible contributions from the ISM CRs) does not vary considerably between SNRs.
This parameter enters into the ratios between IC and Synchrotron emissions to PP emissions.

In section \sref{sec:SNRSab} we applied the results of the earlier sections to show that the broad-band spectra of the SNRs, \SNRa{} and \SNRb{} are inconsistent with a PP origin and are consistent with an IC origin of the $\gtrsim$ TeV emission.
A PP dominated $\TeV$ emission would imply radio synchrotron and probably thermal X-ray Bremsstrahlung fluxes that would greatly exceed the observed X-ray flux.

The neutrino flux from these SNRs is expected to be lower than $\vep_{\nu}f_{\vep_{\nu}}\lesssim10^{-12}\erg\cm^{-2}\se^{-1}$ and is probably too low to be detected by current and planned neutrino observatories.

We compared our main results with previous studies of these SNRs (Tables \ref{table:ICtoPP} and \ref{table:PPtoSyn}) and showed that our simple analytical expressions are in good agreement with more detailed calculations.
All models, in which the $\gamma$-ray emission is dominated by PP, avoided the implied excessive synchrotron emission (but not the implied excessive thermal X-ray Bremsstrahlung emission, see \sref{sec:SNRsabPP}) by assuming an extremely low value of $n^{-1}K_{ep}$, $n^{-1}K_{ep}\lesssim 10^{-3}$.
Such low values of $n^{-1}K_{ep}$ are not plausible since a high density $n\gg0.1$ is inconsistent with the lack of observed thermal X-ray emission and a low value of $K_{ep}\lesssim10^{-4}$ is inconsistent with radio observations of SNRs in nearby galaxies as shown in \sref{sec:M33}. Previous claims, that the $\gamma$-ray emission in SNRs \SNRa{} and \SNRb{} is not IC where discussed in \S~\ref{sec:ClaimsNotIC}.

Interpretation of the narrow filaments seen in the X-ray pictures as cooling width of the emitting electrons was used to obtain magnetic field estimates of order $\sim 100 \muG$ \citep{Berezhko06,Volk05,Bamba05b}. This would rule out an IC source and thus seem implausible. As an illustration, this would require a value of the electron:proton ratio of $K_{ep}\sim 10^{-5}n_{-1}B_{-4}^{-3/2}$ to explain the $\sim 100$ ratio of $\TeV$ to $\GHz$ fluxes per logarithmic frequency (without solving the thermal-Bremsstrahlung problem).
The interpretation of the narrow filaments as cooling width of the multi-TeV X-ray emitting electrons implies that similar filaments are not expected in the radio observations.
There are at least two examples (Tycho's SNR and the remnant of SN1006) where similar filaments are observed in both radio and X-rays. This puts into question the high $B$ interpretation of the X-ray filaments (see discussion in \sref{sec:ClaimsNotIC}).

We note that the Synchrotron to PP ratios would be affected if the magnetic field is enhanced in a small region behind the shock as suggested above but that the conclusion that a PP model requires low values of $K_{ep}$ would not change. To see the effect of thin enhancement regions, assume an extreme case where there is a strong magnetic field $B$ in a small region $d\ll R$ behind the shock, and a negligible magnetic field elsewhere. Assuming that the accelerated electrons are not confined to this region, the radio emission would be proportional to $dB^{3/2}$ and it would be possible to allow for a higher value of $K_{ep}$ in a PP model for a given value of the magnetic field. Note however, that in order to cool the electrons emitting the $\TeV$ IC for a given SNR age (see discussion in \sref{sec:ICSupp}), the magnetic field would have to be larger in order to cool the electrons in the time they reside in the high magnetic field region and thus will have to be larger by a factor $\propto d^{-1/2}$ compared to a homogenous case. So the suppression of the radio flux due to the small emitting region, given that the IC emitting electrons are cooled, will be roughly equal to $(d/R)^{-1/4}$ where $R$ is the remnant radius. The thin filaments observed by chandra, have widths of $~2'$ and $~1'$ for \SNRa{} and \SNRb{} respectively \citep{Berezhko06,Volk05,Bamba05b}. Taking into account a projection factor of $\approx 7$ \citep{Berezhko06} the emission regions widths are fractions $d/R\sim 10^{-2}$ and $3\X10^{-3}$ of the SNR radii respectively. This would require a correction factor of $(d/R)^{1/4}\sim 3-5$ to equation Eq. \eqref{eq:PPSynICSuppSNRsab} and will not change the conclusions. Furthermore, assuming that the X-ray synchrotron emitting electrons are also effectively cooled in this region, Eq. \eqref{eq:PPSynXICSuppSNRsab} will remain valid.

Using the magnetic field value $B\sim 10\muG$, the ratio of magnetic field energy to thermal energy of swept up material is roughly given by:
\begin{align}\label{eq:epBGeneral}
\ep_B\sim 3\frac{B^2}{8\pi}R^3E_{\text{swept}}^{-1}\sim 4\X10^{-4} B_{-5}^2R_1^3E_{\text{swept},51}^{-1},
\end{align}
where $E_{\text{swept}}=10^{51}E_{\text{swept},51}\erg$ is the total energy in swept up material. We note that if the distances to these SNRs are a few kpcs
$\ep_B$ would equal a few percents. For example, a radius of $R=30R_{1.5}\pc$, implying distances of $3R_{1.5}\kpc$ and $1.5R_{1.5}\kpc$ to \SNRa{} and \SNRb{} respectively, implies $\ep_B\sim 0.01B_{-5}^2R_{1.5}^3E_{\text{swept},51}^{-1}$ and is consistent with all observations.
We note that larger distances imply smaller densities since the velocity is limited from below by Eq. \eqref{eq:MinVsNTXSNRsab}, $\vv_s>3\X10^{8}\nukeV^{1/2}\cm\se^{-1}$, and the number density can roughly be expressed as $n\sim 2\X10^{-3} E_{\text{swept},51}\vv_{8.5}^{-2}m_p^{-1}R_{1.5}^{-3}\cm^{-3}$ where $\vs=3000\vv_{8.5}\km~\se^{-1}$.
Such low densities are expected if these shocks are propagating into progenitor winds \citep[see e.g. ][ and refferences within]{Berezhko06}.

$\gamma$-ray observations in the GeV to sub-TeV range by the GLAST experiment will hopefully allow a clear direct distinction between the IC predicted spectrum, $\nu f_\nu\propto \nu^{1/2}$ (which is thus predicted for the SNRs \SNRa{} and \SNRb{}) and the PP predicted spectrum $\nu f_\nu\propto \nu^{0}$ (with a cutoff at $\sim 100 \MeV$ energies).
Using Eq.~\eqref{eq:ICPP}, the expected IC to PP flux ratio for \GeV{} photon energies is approximately
\begin{align}\label{eq:ICPPGeV}
L_{\nu~\IC}/L_{\nu~\PP}(\GeV)\approx 3K_{ep,-2}n_{-1}^{-1}.
\end{align}
A non-negligible contribution of the PP emission cannot be ruled out (for smaller photon energies the PP emission is strongly suppressed). However, it is certainly possible that PP emission is masked out by IC at all photon energies for these SNRs.

An interesting question is what kind of SNR parameters are required in order to have an observable gamma ray emission dominated by PP collisions. Higher densities would result in higher PP emission albeit with lower maximal proton energy.
The maximal proton energy is proportional to ${\vep_{p,\text{max}}\propto BR\vv_s\propto EBR^{-1/2}n^{-1/2}}$.
Based on the observation that electrons are accelerated to $\sim 60 \TeV$ energies in these SNRs we assume protons are accelerated to similar energies (probably somewhat higher if the electrons are limited by cooling). Therefore, comparing to these SNRs, we have freedom to increase the density by a factor of $\sim 100$ (fixing the energy, radius and magnetic field), while keeping protons energetic enough to produce $\sim 1\TeV$ photons.
The cutoff photon energies in the IC spectrum and the synchrotron spectrum are both proportional to
$\propto \vv_s^2B^{-1}\propto ER^{-3}B^{-1}n^{-1}$ and $\propto \vv_s^2B^2R^2\propto ER^{-1}B^2n^{-1}$, for cooling and age limits respectively [see Eqs. \eqref{eq:hnuSynST} and \eqref{eq:hnuICST}]. A factor of $\sim 100$ in the density (for fixed energy, radius and magnetic field) would shift the IC and synchrotron cutoff energies by a factor of $1/100$, strongly suppressing the $\TeV$ IC and $\keV$ synchrotron emissions.
At the same time, a larger density would increase the thermal X-ray emission (as long as the post shock temperature does not fall below the X-ray observable energies). It is therefore likely that SNRs with considerably higher ambient densities have observable PP dominated $\TeV$ emission. Such SNRs will have thermal or no observable X-ray radiation rather than non-thermal X-ray radiation.
At $\sim 1\GeV$ photon energies, densities exceeding $n\gtrsim0.3 K_{ep,-2}^{-1}$ [cf. \eqref{eq:ICPPGeV}] are enough for PP emission to dominate the IC emission.

Neutrino emission from PP collisions is similarly expected to be higher in SNRs evolving in high density environments (the neutrino flux roughly equals the PP gamma ray flux) and are likely to be better observed in SNRs with strong thermal X-ray emission (or no X-ray emission).
For SNRs with observed thermal X-ray emission, the expected neutrino flux can be estimated directly using \eqref{eq:PPTB2}.

We conclude that there is need for a detailed analysis using the X-ray and radio data of SNRs in order to find suitable candidates for PP $\gamma$-ray and neutrino emission. The analytical tools developed in this paper may be used to estimate the expected $\gamma$-ray and neutrino fluxes and to determine the dominant $\gamma$-ray emission process based on existing radio and X-ray observations of SNRs.

\acknowledgements We thank M. Fukugita for discussions that triggered this work. This research was partially supported by ISF, AEC and Minerva grants.

\appendix
\section{Emission mechanisms}\label{sec:EmmisionsA}
\subsection{Thermal Bremsstrahlung}
The thermal Bremsstrahlung emissivity per unit frequency of an optically thin plasma with temperature $T_e$ is given by \citep{Rybicki79}:
\begin{align}
&\ep_\nu^{ff}=\frac{2^5\pi q^6}{3m_ec^3}\left(\frac{2\pi}{3m_e}\right)^{1/2}T_e^{-1/2}Z^2n_en_ie^{-h\nu/T_e}\bar g_{ff},
\end{align}
where $n_e,~T_e$ are the electron number density and temperature respectively, $n_i,~Z$ are the ions' number density and charge respectively, and $\bar g_{ff}$ is the thermal Gaunt factor.
For a plasma consisting of electrons and protons with equal number density $n$ we have
\begin{align}
&\nu\ep_\nu^{ff}=\sqrt{\frac{8}{3\pi}}\sig_T\al_ec\left(\frac{m_ec^2}{T_e}\right)^{1/2}n\frac{h\nu}{T_e}T_ene^{-h\nu/T_e}\bar g_{ff}.
\end{align}

The function $xe^{-x}$ attains its maximal value ($e^{-1}$) at $x=1$. The maximal Luminosity per logarithmic frequency is thus
\begin{align}\label{eq:TBExactA}
&\nu L_{\nu,h\nu=T_e}^{ff}=\sqrt{\frac{8}{3\pi}}e^{-1}\al_e\bar g_{ff}N\sig_Tnc\sqrt{m_ec^2T_e}^{1/2}.
\end{align}
For $h\nu=T_e$, $100\eV<T_e<10\keV$, the value of $\bar g_{ff}$ is in the range, $0.8<\bar g_{ff}<1.2$ \citep[e.g. ][]{Karzas61}.

We next consider the expected value of the electron temperature due to coulomb heating by protons.
The equation for the change in the electron temperature due to coulomb collisions with protons is given by \citep[eg. ][]{Ichimaru04}:
\begin{equation}
\frac{dT_e}{dt}=(T_p-T_e)\frac{8\sqrt{2\pi}nq^4}{3m_em_p}\left(\frac{T_e}{m_e}+\frac{T_p}{m_p}\right)^{-3/2}\lm_{ep},
\end{equation}
where $\lm_{ep}$ is the Coulomb logarithm. Assuming that $m_e/m_p\ll T_e/T_p\ll1$ the electron temperature after a time $t=t_{\text{kyr}}\kyr$ will be:
\begin{align}\label{eq:TeTpCoulomba}
 &T_e\sim 0.6 \left(\lm_{ep,1.5}n_0t_{\text{kyr}}T_{p,\text{keV}}
\right)^{2/5} \keV,
\end{align}
where $\lm_{ep}=30\lm_{ep,1.5}$. For $10^{-2}<T_e/T_p<0.6$ the correction to this expression is smaller than $20\%$.

Next consider power law distributions of accelerated electrons or protons,
\begin{align}
\frac{dN_i}{d\gamma_i}=A_i\gamma_i^{-p},
\end{align}
where $i=e,p$.

\subsection{Gamma rays from proton-proton collisions}
The spectrum of emitted photons is given by \cite{Drury94} :
\begin{align}
&\inv{h}L_{\nu~\PP}=\ave{mx}^p_{\gamma}\sig_{pp}^{\text{inel}}nc\frac{dN_p}{d\vep_p}|_{h\nu},\cr
&\ave{mx}^p_{\gamma}\approx\frac{2}{p}\ave{mx}^p_{\pi_0}, \end{align}
where $\sig_{pp}^{\text{inel}}$ is the inelastic proton-proton cross section and $\ave{mx}^p_{S}$ is the spectrum weighted moment for particles of type $S$.
This can be written as:
\begin{align}\label{eq:PPExactA}
&\nu L_{\nu~\PP}=C_{\PP}(p)2\vep_p\frac{dN_p}{d\vep_p}\sig^{\text{inel}}_{pp} n c h\nu,
\end{align}
where $\vep_p dN_p/d\vep_p$ is to be evaluated at $\vep_p(\nu)=10h\nu$. Ignoring the correction factor $C_{\PP}(p)$, this is equivalent to assuming that each inelastic p-p collision produces two photons with energy
$h\nu=\vep_p/10$. The correction factor is given by
\begin{align}
&C_{\PP}(p)=\frac{\frac{2}{p}\ave{mx}^p_{\pi_0}}{2\X10^{-(p-1)}}.
\end{align}
For $p=2,2.2$ we have $C_{\PP}(2)\approx 0.85, C_{\PP}(2.2)\approx 0.66$ \citep[values of $\ave{mx}^p_{\pi_0}$ were taken from ][]{Drury94}.

\subsection{IC radiaion of CMB photons}
The spectrum of IC scattered photons of a black body target with temperature $T$ in the Thompson regime is given by \citep{Rybicki79}
\begin{align}
\inv{h}L_{\nu~\IC}=A_e\frac{8\pi^2r_e^2}{h^3c^2}(T)^{(p+5)/2}F(p)(h\nu)^{-(p-1)/2},
\end{align}
where
\begin{align}
F(p)=2^{p+3}\frac{p^2+4p+11}{(p+3)^2(p+5)(p+1)}\Gamma\left(\frac{p+5}2\right)\zt\left(\frac{p+5}2\right).
\end{align}
This can be written as,
\begin{align}\label{eq:ICExactA}
\nu L_{\nu~\IC}=C_{\IC}(p)\haf \vep_e(\nu)\frac{dN_e}{d\vep_e}\frac43 \sig_T\gamma_e^2(\nu)U_{T}c,
\end{align}
where $\vep_edN_e/d\vep_e$ is to be evaluated at $\vep_e(\nu)=\gamma_e(\nu)m_ec^2\equiv m_ec^2(h\nu/3T)^{1/2}$ and $U_T=aT^4$ is the energy density in the black body photons. Ignoring the correction factor $C_{\IC}(p)$, this is equivalent to assuming that each electron emits all the power $\frac43\sig_T\gamma^2U_{T}c$, in photons
of energy $h\nu=\gamma^23T$. The correction factor is given by:
\begin{align}
&C_{\IC}(p)=3^{-(p-7)/2}\frac{15}{16}\pi^{-4}F(p).
\end{align}
For $2<p<2.2$ we have $C_{\IC}(p)\approx 0.8$ (to within $5\%$).
It is useful to note that $\gamma_e^2(\nu)U_{T}=[U_T/(3T)]\nu\approx 0.9 n_T h\nu$ where $n_T$ is the number density of black body photons.

\subsection{Synchrotron radiation}
The spectrum of synchrotron radiation is \citep{Rybicki79}
\begin{align}
&L_\nu=\inv{p+1}\Ga\left(\frac p4+\frac{19}{12}\right)\Ga\left(\frac p4-\inv{12}\right)\frac{\sqrt{3}q^3B\sin\al}{mc^2}
\left(\frac{2\pi mc\nu}{3qB\sin\al}\right)^{-(p-1)/2},
\end{align}
where $\alpha$ is the angle between the electrons velocity and the magnetic field direction.
Using,
\begin{align}
&\inv{4\pi}\int_0^\pi 2\pi\sin\al d\al (\sin\al)^{\frac{p+1}2}=\frac{\sqrt{\pi}\Ga(\frac{p+5}4)}{2\Ga(\frac{p+7}4)},
\end{align}
the spectrum for an isotropic distribution is
\begin{align}
&L_\nu=A_eD(p)\frac{q^3B}{mc^2} \left(\frac{mc\nu}{qB}\right)^{-(p-1)/2}
\end{align}
with
\begin{align}
D(p)=\inv{p+1}\frac{\sqrt{\pi}\Ga\left(\frac{p+5}4\right)\Ga\left(\frac p4+\frac{19}{12}\right)\Ga\left(\frac
p4-\inv{12}\right)}{2\Ga\left(\frac{p+7}4\right)}\sqrt{3}\left(\frac{2\pi}{3}\right)^{-(p-1)/2}.
\end{align}
This can be written as
\begin{align}\label{eq:SynExactA}
&\nu L_\nu=C_{\Syn}(p)\haf \vep_e(\nu)\frac{dN_e}{d\vep_e}\frac43\sig_T\gamma_e^2(\nu)U_Bc,
\end{align}
where $\vep_e~dN_e/\vep_e$ is to be evaluated at $\vep_e(\nu)=\gamma_e(\nu)m_ec^2\equiv (2\nu/\nu_B)^{1/2}m_ec^2$, $\nu_{B}\equiv qB/(2\pi m_ec)$ and $U_{B}=B^2/(8\pi)$.
Ignoring the correction factor $C_{\Syn}(p)$, this is equivalent to assuming that each electron emits all it's power of $(4/3)\sig_T\gamma^2U_{B}c$ in photons of energy $h\nu=\gamma^2h\nu_{B}/2$. The correction factor is given by
\begin{align}
&C_{\Syn}(p)=\frac{9}{2}\X(4\pi)^{\frac{p-3}{2}}D(p)
\end{align}
and is approximately $C_{\Syn}(p)\approx 0.8$ (to within $5\%$) for $2\leq p<2.2$.

\bibliographystyle{apj}

\end{document}